\documentclass[a4paper,11pt]{article}
\pdfoutput=1
\usepackage{jheppub}
\usepackage[T1]{fontenc} 
\usepackage{amsmath}
\usepackage{amsfonts}
\usepackage{amssymb}
\usepackage{mathrsfs}
\usepackage{mathtools}
\usepackage[english]{babel}

\title{\begin{center}
\boldmath Tensor Adiabatic Modes and Consistency Relations\\ with Primordial Axion-Gauge Fields
\end{center}
}
\author{Azadeh Maleknejad}

\affiliation{School of Physics, Institute for Research in Fundamental Sciences (IPM),\\ P. Code. 19538-33511, Tehran, Iran}
\emailAdd{azade@ipm.ir}

\abstract{We study the tensor consistency relation in models of axion inflation with an SU(2) gauge field. In the tensor sector, we have two spin-2 modes, the standard gravity waves and the tensor perturbations of the SU(2) gauge field which sources the gravity waves at the linear level. Interestingly enough, we find that the gravity waves are adiabatic and Maldacena's consistency relation including a long wavelength gravity wave holds in this setup. However, since it is partially polarized, there is a difference between the (n+1)-point functions with different helicity states proportional to the ratio of the gauge field density to the total density. These chiral n-point functions are the imprints of the SU(2) gauge field on the primordial cosmological perturbations and a robust observational feature of their contribution to the physics of inflation.}

\usepackage{amsmath}
\usepackage{amsfonts}
\usepackage{amssymb}
\usepackage{mathrsfs}
\usepackage{mathtools}
\usepackage[english]{babel}
\usepackage{myaasmacros}
\usepackage{iopams}
\usepackage{tensor}
\usepackage{graphicx}
\usepackage{bm}
\usepackage[dvipsnames]{xcolor}
\usepackage{xcolor,colortbl}
\usepackage{color}
\usepackage{mathrsfs}
\usepackage{mathtools}
\usepackage{slashed}
\usepackage{hyperref}
\usepackage{tikz}

\def \dd  {{\rm d}}

\newcommand{\bse}{\begin{subequations}}
\newcommand{\ese}{\end{subequations}}
\newcommand{\mpl}{M_{pl}}
\newcommand{\e}{\textbf{e}}

\newcommand{\be}{\begin{equation}}
\newcommand{\ee}{\end{equation}}
\def\bea{\begin{eqnarray}}
\def\eea{\end{eqnarray}}

\usepackage{xcolor,colortbl}
\definecolor{darkred}{rgb}{0.7,0,0}
\definecolor{deepcerise}{rgb}{0.85, 0.2, 0.53}

\def\dre_g{\delta\rho_g}
\def\dpe_g{\delta P_g}
\def\dqe_g{\delta q_g}
\def\dre{\delta\rho}
\def\dpe{\delta P}
\def\dqe{\delta q}

\def\mH{\mathcal{H}}

\def\YM1{\frac{\dot\phi^2}{a^2}}
\def\YM2{\frac{g^2\phi^4}{a^4}}

\def\dd{\delta\psi}
\def\y{\tilde{x}}

\def\th{\tilde{h}}
\def\tg{\tilde{\gamma}}

\definecolor{blue-green}{rgb}{0.0, 0.87, 0.87}

\newcommand{\dt}{\delta\!_{_{\rm T}}}

\newcommand{\dtwo}{\delta\!_{_{2}}\!}
\newcommand{\g}{\textrm{g}}

\begin{document}

\maketitle

\flushbottom

\section{ Introduction}

The energy scale of inflation can be as high as $10^{14}$ GeV which is perhaps the highest observable energy scale in the Universe.
Therefore it is natural to tune the inflationary model within the existing particle physics models suitable for similar energy scales.
Axion fields are abundant in string theory and therefore well-motivated candidates
for the inflaton field. The axion effective potential is protected from
dangerous quantum corrections, thanks to the shift symmetry which guaranteed the flatness of the potential. The axion
field, $\varphi$, is classically coupled to gauge fields through the Chern-Pontryagin density, $F\wedge F$. In fact, gauge fields are commonplace in all particle physics, and, in particular, beyond standard models which could have contributed to the physics underlying inflation. Consequently, one important question is the possible observational features that the primordial axion-gauge fields may leave on the CMB and LSS which hints their contribution to the early Universe physics. For an extensive review on the models of axion inflation and gauge fields in the physics of inflation see \cite{Maleknejad:2012fw} and \cite{Pajer:2013fsa}.

The squeezed limit $n$-point functions are perhaps the easiest to measure, and besides, carriers very direct information about the spectrum of particles during inflation \cite{Arkani-Hamed:2015bza}. The inflationary consistency relation is a powerful probe of the early universe physics which holds under very general conditions, i.e. when the long-wavelength mode is adiabatic \cite{Maldacena:2002vr}. 
 The key points for deriving the consistency relation are the Weinberg's adiabatic modes \cite{Weinberg:2003sw}, which are constant at $k/\mH\ll1$, and can be removed by a local coordinate transformation of the background metric. The conservation of these adiabatic modes at super-horizon scales is essential for relating the cosmological fluctuations produced during inflation with those observed in CMB and LSS. The Weinberg's theorem states that: assuming the standard Banch-Davis initial condition, regardless of the details, inflationary settings with vanishing entropy perturbations and anisotropic stress at super-horizon scales always have two scalar and two tensor adiabatic solutions \cite{Weinberg:2003sw}.

 The consistency relation connects the squeezed limit $(n+1)$-point function involving a long adiabatic mode and $n$ shorter modes to the $n$-point function of the shorter modes. Maldacena's original consistency relation has been generalized to the conformal consistency relation in \cite{Creminelli:2012ed} and to an infinite set of Ward identities in \cite{Hinterbichler:2013dpa}.  Assuming Banch-Davis vacuum, the scalar consistency relation holds in single clock inflationary models which are dynamical attractors \cite{Creminelli:2004yq}. \footnote{Non-attractor single-field models can generate large local non-Gaussianities and violate the consistency relation \cite{Chen:2013aj, Flauger:2013hra}.} On the other hand, Multi-field scalar models of inflation typically violates the scalar consistency relations. However, as far as the multi-field model has vanishing anisotropic stress for $-k\tau\ll1$, the gravity wave consistency relation holds \cite{Cosmology-W, Dimastrogiovanni:2015pla}. The inflationary extended cosmic no-hair theorem, \footnote{First argued by Hawking, Gibbons, and Moss in \cite{PhysRevD.15.2738, HAWKING198235}, this statement that the late-time behavior of any accelerating Universe is an isotropic Universe, is called cosmic no-hair conjecture. The first attempt to prove this conjecture was presented in Wald's seminal paper \cite{Wald:1983ky} which states that: Bianchi-type models (except Bianchi IX) with the total energy-momentum tensor made of a cosmological constant, $\Lambda_0$, and $\tilde T_{\mu\nu}$ satisfying strong and dominant energy conditions approach de Sitter space exponentially fast, within a few Hubble times $H^{-1}=\sqrt{3/\Lambda_0}$. However, inflationary setups do not satisfy the conditions of the Wald's theorem \cite{Maleknejad:2012as}. } presented in \cite{Maleknejad:2012as}, states that such inflationary models violate the cosmic no-hair conjecture and hence can have stable anisotropic inflation solutions. Inflation, however, puts an upper bound on the growth of anisotropy which for slow-roll dynamics is of the order of slow-roll parameters. Among models that can violate tensor consistency relation, one can mention the solid inflation \cite{Endlich:2013jia} and anisotropic inflation \cite{Bartolo:2012sd}. The violation of the cosmic no-hair conjecture and possibility of the anisotropic inflation in these models has been studied in \cite{Watanabe:2009ct, Bartolo:2013msa}.
A more recent and model independent discussion on the violation of the tensor consistency relation is presented in \cite{Bordin:2016ruc}. 


The focus of this work is on the single field axion inflation models coupled to an SU(2) gauge field. \footnote{
Among the inflationary models involving an SU(2) gauge field, we can mention gauge-flation and chromo-natural inflation which have been introduced and studied in \cite{Maleknejad:2011jr} and \cite{Adshead:2012kp} respectively. Both of gauge-flation and chromo-natural inflation models have been disfavored by Planck data \cite{Dimastrogiovanni:2012ew, Adshead:2013nka}. Recently, two axion-gauge field inflationary models consistent with the current data have been studied in \cite{Maleknejad:2016qjz} and \cite{Adshead:2016omu}.} The gauge field and the axion are coupled through a Chern-Pontryagin density, $\frac{\lambda}{f}F\wedge F$. Due to the SU(2) algebra, the gauge field can have an isotropic and homogeneous field configuration. The interaction with the axion is then essential for breaking the conformal invariance and evading the $1/a$ decay of the gauge field. We assume that the gauge field has a small VEV and a negligible effect on the background evolution, i.e. $\rho_{_{\rm YM}}\lesssim\epsilon^2\mpl^2H^2$. Moreover, for the sake of generality, we consider an arbitrary axion potential that is able to support the slow-roll inflation. The cosmic perturbations of this model has been recently studied in \cite{Maleknejad:2016qjz} and has been shown that it is in agreement with the current CMB observations. \footnote{ For an intrinsic inflationary leptogenesis mechanism in models of axion inflation with a classical SU(2) gauge field see \cite{Maleknejad:2016dci, Noorbala:2012fh, Maleknejad:2014wsa}. Interestingly, the model presented in \cite{Maleknejad:2016qjz} can generate the observed value of baryon to photon number density in a natural range of its parameters. } Although, the gauge field has a small contribution to the background evolution, its quantum fluctuations, however, makes a significant contribution to the cosmic perturbations. More precisely, the perturbed gauge field has some scalar degrees of freedom, linearly coupled to the curvature perturbations which will be relevant after horizon crossing. As a result, the scalar perturbations are not adiabatic but we have a slow-roll supported entropy perturbation. That then indicates that the scalar consistency relation is violated in the presence of the SU(2) gauge field.
In the tensor sector, the perturbed gauge field has a spin-2 fluctuation which is linearly coupled to the gravity waves and explicitly breaks the parity between the left- and right-handed polarization states. That then leads to a chiral tensor power spectrum which satisfies in a modified version of the Lyth bound. Since in this model, the gravity waves are coupled to the spin-2 perturbations of the gauge field, that is important to explore if the gravity waves are adiabatic or not and the tensor consistency relation holds.

Recalling that the tensor consistency relation holds in most of the inflationary settings, any robust feature in the gravity waves observables can provide a clean hint for primordial axion-gauge fields. In this work, we focus on the behavior of the super-horizon tensor modes and $(n+1)$-point functions including a long-wavelength gravity wave in this setup. First, we contract the gauge-invariant (internal) spin-2 perturbation of the gauge field by removing the space-time perturbations induced on the SU(2) gauge field. Then we show that in terms of that tensor modes and the gravity waves, the diffeomorphism invariance forbids the presence of any effective mass term for the graviton. We emphasize that the above argument is true for any covariant system of SU(2) gauge fields and is not limited to this particular model. Next, we show that the resulting gravity waves are adiabatic and therefore equivalent to a change of coordinate in the background metric. However, since the gravity wave is chiral, the large diffeomorphism required to remove it is a bit different from the standard one. In fact, the long-wavelength inhomogeneous solution of the gravity wave (which is sourced by the gauge field) has the form of a standing circularly polarized gravity wave in real space which looks like a Bianchi type VII$_0$ metric. 
Finally, we turn to the tensor consistency relation involving a long gravity wave mode. We show that Maldacena consistency relation still holds. However, since the gravity waves are partially polarized, there is a difference between the $(n+1)$-point functions of the different helicity states. That provides an interesting observational feature for these models which can be investigated by the future ambitious CMB observations, e.g. CMB-S4 \cite{Abazajian:2016yjj}. 
We emphasize that since the tensor perturbations in gauge-flation, chromo-natural inflation, and our model are the same, our results hold for those models as well.


The outline of the paper is as follows: In Section \ref{review} we review the symmetries of adiabatic
fluctuations and consistency relations in FRW cosmologies. In section \ref{setup} we briefly review the axion-gauge field inflationary setup. In section \ref{tensor-eq-sol} we investigate the implications of the diffeomorphism invariance on the tensor perturbations and present the solutions. In section \ref{soft} we study the adiabatic nature of the gravity waves in the presence of the spin-2 perturbations of the gauge field and work out the tensor consistency relation that holds in this setup. We conclude in section \ref{conclude}. Some of the technical details of the derivations are provided in Appendices \ref{gauge-invariant}, \ref{app-action} and \ref{Bianchi-app}. Here and throughout, the reduced Planck mass is set to unity, unless otherwise specified.

\section{Adiabatic modes and inflationary consistency relations}\label{review}

Here, we briefly review Weinberg's adiabatic modes theorem and Maldacena's inflationary consistency relations. 
For more details on Weinberg's adiabatic modes see \cite{Weinberg:2003sw, Cosmology-W} and for details on Maldacena's inflationary consistency relations see the seminal paper \cite{Maldacena:2002vr}. The adiabatic modes and consistency relation can be extended to include
the gradient expansions. Maldacena's original consistency relation has been generalized to the conformal consistency relation in \cite{Creminelli:2012ed} and to an infinite set of Ward identities in \cite{Hinterbichler:2013dpa}. 

\subsection{Weinberg's Adiabatic modes}

We start with the construction of Weinberg's adiabatic modes in the Newtonian gauge which is essential for the derivation of the consistency relations. Fixing the gauge, uniquely specifies all the modes with finite momentum, $k\neq0$. However, there are residual gauge symmetries for the very long-wavelength modes, which remain a symmetry of the gauge-fixed action. Starting with the flat FRW metric as the unperturbed metric
\be
ds^2=-dt^2+a^2d\textbf{x}^2,
\ee
here we focus on the implications of the diffeomorphism invariance on the nature of the very long-wavelength modes with $k/\mH\ll1$. Therefore, only the homogeneous diffeomorphisms are relevant. 
The general spatially homogeneous perturbed metric in the Newtonian gauge is
\be
ds^2=-(1+2\Phi(t))dt^2+a^2\bigg((1-2\Psi(t))\delta_{ij}+\gamma_{ij}(t)\bigg)dx^idx^j,
\ee
where $\Phi$ and $\Psi$ are the Bardeen potentials and $\gamma_{ij}$ is the traceless tensor perturbation, gravity wave. \footnote{Here we neglect the vector perturbations due to their damping nature in most of the inflationary models including our axion-gauge field model.} One can decompose the general spatially homogeneous perturbed energy-momentum in the Newtonian gauge as
\bse
\begin{align}
T_{00}&=-\bar{\rho}g_{00}+\delta\rho(t) \quad \textmd{and} \quad T_{0i}=-(\bar{\rho}+\bar{P})\partial_i\delta u(t),\\\label{Tij}
T_{ij}&=\bar{P}(t)g_{ij}(t)+a^2\bigg(\delta_{ij} \delta P(t)+\partial_{ij}^2\pi^S(t)+\pi^T_{ij}(t)\bigg),
\end{align}
\ese
where a bar denotes an unperturbed quantity, $\delta \rho$, $\delta P$ and $\delta u$ are the perturbed density, pressure and velocity potential respectively. Moreover, $\pi^S$ and $\pi^T_{ij}$ are the scalar and tensor anisotropic inertia which characterize departures of $T_{\mu\nu}$ from the perfect fluid form. \footnote{It is interesting to note that in the decomposition \eqref{Tij}, the effects of bulk viscosity are included in $\delta p$ \cite{Weinberg:2003sw}. }

Under the action of diffeomorphism transformations 
\be
x^\mu\mapsto \tilde{x}^{\mu}=x^\mu+\epsilon^\mu(t,\textbf{x}),
\ee
there is a $\epsilon^\mu$ which generates spatially homogeneous transformations on the metric and preserve the Newtonian gauge  \cite{Cosmology-W}
\bse
\begin{align}
&\epsilon^0(t,\textbf{x})=-f(t)-\chi(\textbf{x}),\\
&\epsilon^i(t,\textbf{x})=(\theta\delta_{ij}+\sigma_{ij})x^j-\partial_i\chi(\textbf{x})\int \frac{dt}{a^2(t)},
\end{align}
\ese
where $\theta$ is a constant scalar and $\sigma_{ij}$ is a constant, traceless and symmetric matrix\footnote{In general, $\epsilon^i(t,\textbf{x})$ can have a constant term $\epsilon^i_0$ as well as a term like $\omega_{ij}x^j$ where $\omega_{ij}=-\omega_{ji}$. Here, however, we ignored them because (due to the spatial translational and rotational symmetry of the background metric) they do not have any effects on the linear perturbed metric. }, $\sigma_{ii}=0$. Therefore, choosing the Newtonian gauge, we are still left with residual gauge symmetries for the zero wavenumber modes. These diffeomorphisms do not vanish at spatial infinity and therefore are called large gauge transformations.  The scalar functions $f(t)$, $\chi(t)$ and $\theta$ act only on the scalar perturbations 
\be
\Phi(t)\mapsto \Phi(t)+\dot{f}(t) \quad \textmd{and} \quad \Psi(t)\mapsto \Psi(t)+\theta-Hf(t),
\ee
 while keep the tensor perturbations untouched. On the other hand, $\sigma_{ij}$ acts only on the gravitational waves as
\be
\gamma_{ij}(t)\mapsto \gamma_{ij}(t)-2\sigma_{ij}.
\ee
Therefore, if $\Phi(t)$, $\Psi(t)$ and $\gamma_{ij}(t)$ are solutions of the spatially homogeneous Einstein equations, their transformed quantities and their differences are also the spatially homogeneous solutions. In particular, in the scalar sector, we have spatially homogeneous solutions of the form
\be\label{phi-g}
\Phi_{\rm A}(t)=-\dot{f}(t) \quad \textmd{and} \quad \Psi_{\rm A}(t)=Hf(t)-\theta,
\ee
which corresponds to a cosmic fluid given as
\be
\delta\rho_{\rm A}(t)=-\dot{\bar{\rho}}f(t), \quad \delta P_{\rm A}(t)=-\dot{\bar{P}}f(t), \quad \delta u_{\rm A}(t)=f(t),
\ee
with a vanishing scalar anisotropy
\be
\pi^S_{\rm A}(t)=0.
\ee
That leads to a constant comoving curvature ($\mathcal{R}\equiv -\Psi+H\delta u$) and curvature perturbation ($\zeta\equiv -\Psi-H\delta\rho/\dot{\bar{\rho}}$)
\be\label{R-g}
\mathcal{R}_{\rm A}(t)=\zeta_{\rm A}(t)=\theta.
\ee
There is also a spatially homogeneous solution in the tensor sector as
\be\label{gamma-g}
\gamma^{\rm A}_{ij}(t)=2\sigma_{ij},
\ee 
with a vanishing tensor anisotropic inertia
\be
\pi^{T{\rm A}}_{ij}(t)=0.
\ee
Up to now, the solutions \eqref{phi-g} and \eqref{gamma-g} are only gauge degrees of freedoms for the $k=0$ mode and the Weinberg's theorem relates them to the physical modes. The essential step in Weinberg's theorem is as follows. In case that the anisotropic inertia $\pi_{ij}(t,\textbf{k})$ and the entropy perturbations, $\frac{\dot{\bar{\rho}}\delta P-\dot{\bar{P}}\delta\rho}{3(\bar{\rho}+\bar{P})^2}$, vanish in the limit $k/\mH\ll1$, the spatially homogeneous solutions are extendible to modes with $k\neq0$. Since solutions with non-zero wavenumbers have no residual gauge symmetry in the Newtonian gauge, these modes are physical. When solving the linearized Einstein equations, there are two scalar and two tensor physical solutions which are constant at $k/\mH\ll1$ (eq.s \eqref{R-g} and \eqref{gamma-g}), called adiabatic solutions.\footnote{This solution leads to equal values of $\delta\rho_{x}/\bar{\rho}_x$ for all individual elements in the cosmic fluid which explains the name adiabatic.} One immediate consequence of this theorem is that these modes freeze out at horizon crossing and become indistinguishable from a redefinition of the background metric
\be
\bar{g}_{\mu\nu}(t)+\delta g_{\mu\nu}^{\rm A}(t)=\bar{\tilde{g}}_{\mu\nu}(\tilde x),
\ee 
which implies 
\be\label{adiabatic--}
\delta g_{\mu\nu}^{\rm A}(t)=-\mathcal{L}_{\epsilon}\bar{g}_{\mu\nu}(t), 
\ee
where $\mathcal{L}_{\epsilon}$ denotes the Lie derivative with respect to $\epsilon^\mu$.

\subsection{ Maldacena's consistency relations}

Now, we turn to the consistency relations which is a powerful probe of the early universe physics and holds under very general conditions, i.e. when the long-wavelength mode is adiabatic. Therefore, assuming Banch-Davis initial condition, the scalar consistency relation only hold for single clock inflationary models in which the entropy perturbation is zero. However, gravity waves consistency relations hold for more general inflationary models. More precisely, assuming Banch-Davis initial condition, only models in which $\pi^{T}_{ij}\neq0$ at super-horizon scales can violate tensor consistency relations, e.g. solid inflation \cite{Endlich:2013jia} and anisotropic inflation \cite{Bartolo:2012sd}. As shown in \cite{Maleknejad:2012as}, such inflationary models violate the cosmic no-hair conjecture.

Since the tensor modes are the main focus of this work, here we present the consistency relation for the gravity waves. The scalar consistency relation is the same and one only needs to replace $\gamma_{ij}^{\rm A}$ with the (adiabatic) curvature perturbation $\zeta_{\rm A}$. The key physical point behind the consistency relations is the observation that the adiabatic long-wavelength modes can be removed by the local coordinate transformation of the background metric, i.e. \eqref{adiabatic--}. Hence, they act as a classical background for the short wavelength modes, which freeze out much later than the long mode. In particular, an n-point correlation function of the short modes can be written as
\be\label{n-point-cor}
\langle  \zeta(x_1)\zeta(x_2)\cdots\zeta(x_n) \rangle _{\gamma_{ij}^{\rm A}(x)}=\langle \zeta(\tilde{x}_1)\zeta(\tilde{x}_2)\cdots\zeta(\tilde{x}_n) \rangle,
\ee
which Taylor expanding RHS around $x^i$, we find the change of the short distance n-point correlation function as  
\be
\delta\langle \zeta(\tilde{x}_1)\zeta(\tilde{x}_2)\cdots\zeta(\tilde{x}_n) \rangle=\sum_{\rm I=1}^{n}\delta \vec{x}_{\rm I}.\vec{\nabla}_{\rm I}\langle \zeta(x_1)\zeta(x_2)\cdots\zeta(x_n) \rangle +\cdots,
\ee
where $x^i$ and $\tilde x^i$ are related as
\be
\tilde x^i=x^i+\frac12\gamma^{\rm A}_{ij}x^j.
\ee
As a result, the (n+1)-point correlation function including the long-wavelength mode is given as
\be
\langle \gamma_{ij}^{\rm A}(x) \zeta(x_1)\zeta(x_2)\cdots\zeta(x_n) \rangle\simeq\frac12\bigg\langle \gamma_{ij}^{\rm A}(x)\gamma_{kl}^{\rm A}(x)\sum_{\rm I=1}^{n}x^k_{\rm I}\partial_{l}\langle \zeta(x_1)\zeta(x_2)\cdots\zeta(x_n) \rangle \bigg\rangle,
\ee
in which we only keep the dominate term that has the relevant contribution. The above equality is the consistency relation in real space. Going to the Fourier space, we can expand $\gamma_{ij}(\textbf{q})$ as 
\be
\gamma_{ij}(\textbf{q})=\sum_{\lambda=\pm}\gamma_{\lambda}(\textbf{q})\textbf{e}_{ij}(\hat{q},\lambda),
\ee
where $\textbf{e}_{ij}(\hat{q},\lambda)$ are the time-independent polarization tensors, $\textbf{e}_{ij}(\hat q,\lambda)\textbf{e}^{*}_{~ij}(\hat q,\tilde{\lambda})=2\delta^{\lambda\tilde{\lambda}}$, in which $\lambda=\pm$ corresponding to the $\pm2$ helicity states. Moreover, the two-point function at late time is given as 
\be\label{2-point}
\langle \gamma_{\lambda}(\textbf{q})\gamma_{\lambda'}(\tilde{\textbf{q}})\rangle=(2\pi)^{3}\delta^{(3)}(\textbf{q}+\tilde{\textbf{q}})P^{\rm vac}_{\gamma}(q)\delta^{\lambda\lambda'},
\ee
in which 
\be\label{P-vac}
P^{\rm vac}_{\gamma}(q)=q^{-3}\bigg(\frac{H^2}{\mpl^2}\bigg),
\ee
is the power-spectrum.
Then using the above and neglecting the gradients, we arrive at the Maldacena's consistency relation
\be
\langle \gamma_{\lambda}(\textbf{q}) \zeta_{\textbf{k}_1}\zeta_{\textbf{k}_2}\cdots\zeta_{\textbf{k}_n} \rangle'\simeq-\frac12P^{\rm vac}_{\gamma}(q)\sum_{\rm I=1}^n \textbf{e}_{ij}(\hat{q},\lambda) k_{{\rm I}i}\partial_{k_{{\rm I}j}}\langle \zeta_{\textbf{k}_1}\zeta_{\textbf{k}_2}\cdots\zeta_{\textbf{k}_n} \rangle' \quad \textmd{for} \quad q\rightarrow0,
\ee
where the prime in $\langle \cdots \rangle$ indicates that we extracted the prefactor  $(2\pi)^{3}\delta^{(3)}(\textbf{q}+\sum_{\rm I}^n \textbf{k}_{\rm I})$ associated to momentum conservation. Note that the above result follows directly from the fact that adiabatic long-wavelength gravity wave is equivalent to a change of coordinate for the short wavelength mode, regardless of the super-horizon behavior of the short modes. Therefore, as far as our inflationary system generates adiabatic tensor perturbations the above consistency relation holds.

\section{Axion-gauge field inflationary setup}\label{setup}

We consider a generic axion-driven inflation model with a gauge field sector, both minimally coupled to Einstein gravity 
\bea\label{action}
\mathcal{L}_{\textmd{inf}} =\frac{R}{2}-\frac12\partial_\mu\varphi\partial^\mu\varphi-V(\varphi)+\mathcal{L}_{A}(A^a_{\mu},g_{\mu\nu},\varphi)\,,
\eea
where $\varphi$ is the axion field, $V(\varphi)$ is the axion potential and $\mathcal{L}_{A}$ is the gauge field sector.  For the purpose of this work and in order to be as model-independent as possible, $V(\varphi)$ is an arbitrary potential that is able to support the slow-roll inflation. The inflaton then couples directly to the Pontryagin density of a non-Abelian gauge field as
\bea\label{action-A}
\mathcal{L}_{A}(A^a_{\mu},g_{\mu\nu},\varphi)=-\frac{1}{4}\bigg(F^a_{\mu\nu}F_a^{\mu\nu}+\frac{\lambda}{f}\varphi\ F^a_{\mu\nu}\tilde{F}_a^{\mu\nu}\bigg)\,,
\eea
where $\lambda$ is a dimensionless parameter, $f$ is the axion decay constant and $\tilde{F}^{a\mu\nu}=\frac12\epsilon^{\mu\nu\lambda\sigma}F^a_{\lambda\sigma}$. The action \eqref{action} with the standard cosine potential and $\frac{\lambda}{f}\sim\mathcal{O}(10^3)$ is the chromo-natural inflation \cite{Adshead:2012kp, Adshead:2013nka}. The action with the arbitrary potential and $\frac{\lambda}{f}\sim\mathcal{O}(10)$ has been studied in \cite{Maleknejad:2016qjz}.

The field strength tensor is
 \be
 F^a_{~\mu\nu}=\partial_{\mu}A^a_{\nu}-\partial_{\mu}A^a_{\nu}+\g\epsilon^a_{~bc}A^b_{\mu}A^c_{\nu},
 \ee
where $\g$ is the gauge field coupling constant. In the flat FLRW metric
\be
\label{FLRW}
ds^2=-dt^2+a(t)^2\delta_{ij}dx^{i}dx^{j},
\ee
and after choosing the temporal gauge for the gauge field $(A_0=0)$, we have the following isotropic and homogeneous field configuration 
\be\label{ansatz}
A^a_{~0}T_a=0  \quad \textmd{and}\quad A^a_{~\mu}(t)T_a= \psi(t)\bar{e}^a_{~\mu}T_a,
\ee%
where $\{\bar{e}^{\alpha}_{~\mu}\}$ is tetrad field of FRW metric (with $\bar e^{a}_{~0}=0$) 
 satisfying
\be\label{tetrad}
g_{\mu\nu}=e^{\alpha}_{~\mu}e^{\beta}_{~\nu}\eta_{\alpha\beta},
\ee
and $T_a$ are the generators of the SU(2) group
\be
[T_a,T_b]=i\epsilon^{abc}T_c \quad \textmd{and} \quad \textrm{Tr}(T_aT_b)=\frac12\delta^{ab}.
\ee
Here $\alpha,\beta=0,1,2,3$, $a=1,2,3$ and $\eta_{\alpha\beta}$ is the Minkowski metric. 
For the FRW metric, $\{e^{\alpha}_{~\mu}\}$ are specified as 
\be
\bar e^{0}_{~\mu}=\bar n_{\mu} \quad \textmd{and} \quad \bar e^{a}_{~\mu}=a(t)\delta^a_{\mu},
\ee
where $\bar n^{\mu}=(1,0,0,0)$ is the 4-velocity of the comoving observer.
 Note that the effective field value of the gauge field $\psi$ in \eqref{ansatz} is a pseudo-scalar under parity.

The field equations of the unperturbed $\varphi$ and $\psi$ are
\bse\label{c-n.e.o.m}
\begin{align}
&\ddot\varphi+3H\dot\varphi+V_{\varphi}=-3\frac{\lambda \g}{f} \psi^2(\dot\psi+H\psi)\,,\\\label{eq-psi}
&\ddot\psi+3H\dot\psi+(2H^2+\dot H)\psi+2g^2\psi^3=\frac{\lambda \g}{f}\psi^2\dot\varphi\,,
\end{align}
\ese
which are coupled by the Pontryagin density of the gauge field.
For later convenience, we introduce two dimensionless parameters 
\be\label{xi-xi-psi}
\xi\equiv\frac{\lambda\dot{\varphi}}{2fH} \quad \textmd{and} \quad \xi_{\psi}\equiv\frac{\g\psi}{H}.
\ee
 It was showed in \cite{Maleknejad:2016qjz} that a successful inflation in agreement with the current CMB data requires that $\psi\lesssim10^{-2}$, $\sqrt{2}\lesssim\xi_{\psi}<3$ and for a GUT scale inflation with $H\sim10^{-6}\mpl$ and $f\simeq0.1\mpl$, we have $g\sim 10^{-4}$ and $\lambda\sim1$. 

\subsection{Geometry of perturbed gauge field}\label{sec-gauge-invariant}

The metric and SU(2) gauge field are perturbed around their homogeneous and isotropic background configurations (Eqn. \eqref{ansatz}) as
\be\label{field-appdx}
\delta g_{\mu\nu}(t,\textbf{x})=\bar g_{\mu\nu}(t)+\delta g_{\mu\nu}(t,\textbf{x})\quad \textmd{and}\quad A^a_{~\mu}(t,\textbf{x})= \psi(t)\bar{e}^a_{\mu}(t)+\delta A^a_{~\mu}(t,\textbf{x}),
\ee%
where the bar denotes the unperturbed quantity and $\delta A^a_{~\mu}$ involves $3\times4$ components. Therefore, the 12 gauge field perturbations together with the 10 components of the perturbed metric, add up to 22 degrees of freedom. Due to the gauge transformations, not all of that metric and field perturbations are gauge invariant.
In particular, we have two types of gauge freedoms: we call them ``$x^\mu$-gauge'' and ``$A^a$-gauge''.

\vskip 0.3 cm
\noindent \textit{Space-time gauge transformations ($x^\mu$-gauge):} 
\vskip 0.25 cm

Consider a space-time coordinate transformation of the form
\be\label{coord}
x^\mu\mapsto \tilde x^\mu=x^\mu+\epsilon^{\mu}(t,\textbf{x}),
\ee
where $\epsilon^{\mu}$ is small in the sense that it leads to small perturbations in $g_{\mu\nu}$ and $T_{\mu\nu}$.
 Under the above coordinate transformation, the metric tensor and gauge field vector transform as
\be
\tilde g_{\mu\nu}(\y)=g_{\alpha\beta}(x)\frac{\partial x^{\alpha}}{\partial\y^\mu}\frac{\partial x^{\beta}}{\partial\y^\nu} \quad \textmd{and} \quad \tilde A^a_{~\mu}(\y)=A^a_{~\alpha}(x)\frac{\partial x^{\alpha}}{\partial\y^\mu}.
\ee
For our purpose, it is more convenient to work with gauge transformations instead, so-called \textit{$x^\mu$-gauge}, which only acts on perturbations while keeps the unperturbed quantities untouched. Under the action of that gauge transformation, the first order perturbed metric and fields are as follows
\bse\label{Lie-der}
\begin{align}
\delta g_{\mu\nu}&\mapsto \delta g_{\mu\nu}-\mathcal{L}_{\epsilon}g_{\mu\nu}=\delta g_{\mu\nu}-\delta t\dot{\bar{g}}_{\mu\nu}-2\bar{g}_{\lambda(\nu}\partial_{\mu)}\delta x^{\lambda},\\\label{gauge-trans}
\delta A^a_{~\mu}&\mapsto \delta A^a_{~\mu}-\mathcal{L}_{\epsilon} A^a_{~\mu}=\delta A^a_{~\mu}-\dot{\psi}\bar{e}^a_{~\mu}\delta t-\psi\mathcal{L}_{\epsilon}e^a_{~\mu},
\end{align}
\ese
where $\mathcal{L}_{\epsilon}$ is the Lie derivative with respect to $\epsilon^{\mu}$. Moreover, the orthonormal tetrad $\{e_{\mu}^\alpha\}$ transforms as
\be\label{de}
\delta e^{\alpha}_{~\mu}\mapsto \delta e^{\alpha}_{~\mu}-\mathcal{L}_{\epsilon}e^{\alpha}_{~\mu}.
\ee
We realize that the space-time gauge transformations affect the perturbed gauge field and generates a term in $\delta A^a_{~\mu}$ which transforms like $\psi\delta e^a_{\mu}$. We call this geometry induced term $\delta_{\rm x} A^a_{~\mu}$ which is \footnote{Notice that we dropped the slow-roll suppressed term $\dot{\psi}\bar{e}^a_{~\mu}\delta t$ ($\frac{\dot{\psi}}{H\psi}\lesssim\epsilon$).}
\be
\delta_{\rm x} A^a_{~\mu}=\psi(t)\delta e^a_{\mu},
\ee
and we label the rest of $\delta A^a_{\mu}$ which is invariant under $x^{\mu}$-gauge, $\delta\!_{\rm g} A^a_{~\mu}$.
In other words, we can decompose $\delta A^a_{\mu}$ as
\be\label{decompose-A}
\delta A^a_{~\mu}=\delta_{\rm x} A^a_{~\mu}+\delta\!_{\rm g} A^a_{~\mu},
\ee
where $\delta\!_{\rm g} A^a_{~\mu}$ is the genuine gauge field fluctuation which only changes under internal gauge field transformation.

\vskip 0.3cm
\noindent \textit{Internal gauge transformations ($A^a$-gauge):} 
\vskip 0.25cm

Under the action of a an element $U$ of $SU(2)$ near the unit element
\be
U(\lambda^a)=\exp(-\lambda^a(t,\textbf{x})T_a),
\ee  
in which $\lambda^a$ is the infinitesimal gauge transformation parameter, the genuine part of the gauge field transforms as
\be \label{gua}
\delta_{\rm g} A^a_{~\mu}\mapsto \delta_{\rm g} A^a_{~\mu}+\frac{1}{\g}(D_\mu\lambda)^a, 
\ee
where $D_{\mu}=\partial_{\mu}-i\g A^a_{\mu}T_a$ is the covariant derivative. Note that $\delta_{\rm x} A^a_{~\mu}$ which is induced by the geometry is invariant under the gauge field $SU(2)$ transformations. We call the infinitesimal \textit{internal gauge field} transformation \textit{$A^a$-gauge}. 

Using \eqref{decompose-A}, we can decompose the 12 independent components in $\delta A^a_{~\mu}$ as below
\bse \label{gauge-field-pert}
\begin{align}\label{gauge-field-perti}
\delta A^a_{~i}&=a\delta^a_i \delta\psi+\delta^{aj}\big(\partial_{ij}\tilde{Z}+a\partial_i v_j+a\tg_{ij}\big)+a\psi\epsilon^{a~j}_{~i}\big(\g\partial_{j}(\tilde{Z}-Z)+aw_j\big)+\psi \delta e^a_{~i},\\
\delta A^a_{~0}&=\delta^{k}_a\partial_kY+a\delta_a^j u_j+\psi \delta e^a_{~0},
\end{align}
\ese
where $\{\dd, Y, \tilde Z, Z\}$ parametrize scalar perturbations, $\{u_i, v_i, w_i\}$ are divergence-free vector perturbations
and $\tg_{ij}$ is the symmetric, traceless and divergence-free tensor fluctuation. The explicit form of the $\delta e^a_{\mu}$ in the above which induced by space-time geometry is presented in \eqref{tetrad-pert-Appn}. The fields $\{\dd, Y, \tilde Z, Z, u_i, v_i, w_i, \tg_{ij}\}$ are invariant under the action of space-time gauge transformation and therefore, the genuine gauge field fluctuations. However, under the SU(2) gauge transformations, they transform like \eqref{gua}. The gauge transformation parameter $\lambda^a(t,\textbf{x})$ can be decomposed as
$$\lambda^a=\delta^{ai}\partial_i\lambda+\delta^a_i\lambda^{~i}_{V},$$
in which $\lambda$ is the scalar and $\lambda^V_i$ is the divergence-free vector part.
Therefore, among these 12 independent components, one scalar, and one vector mode are gauge degrees of freedom. In appendix \ref{gauge-invariant}, we remove the remaining 3 gauge degrees of freedom and construct the gauge-invariant combinations coming from $\delta A^a_{\mu}$. As we will show in the next section, the spin-2 perturbation $\tg_{ij}$, is invariant under both $x^\mu$-gauge and $A^a$-gauge transformations.

\section{Tensor modes, implications of diffeomorphism invariance}\label{tensor-eq-sol}

In this work, we are interested in the behavior of tensor modes in the presence of the non-Abelian gauge fields. Therefore, in the following, we focus on the tensor fluctuations of the perturbed metric and gauge field. For the thorough study of cosmic perturbations in these models, we refer the interested reader to \cite{ Maleknejad:2016qjz}.

As we showed in the previous section, $\delta A^a_{~i}(t,\textbf{x})$ is not diffeomorphism invariant, but $\delta_{\rm g} A^a_{~i}(t,\textbf{x})$. For instance, under the action of a large, anisotropic transformation of the form
\be\label{large-x}
x^i\rightarrow \tilde x^i=x^i+\sigma_{ij}x^j,
\ee
where $\sigma_{ij}$ is a traceless, the tensor perturbations of the metric and gauge field transform as (see \eqref{Lie-der})
\bea\label{g-trans}
&&\gamma_{ij}(t,\textbf{x})\mapsto \gamma_{ij}(t,\textbf{x})-2\sigma_{ij},\\
&&\dt A^a_{~i}(t,\textbf{x})\mapsto \dt A^a_{~i}(t,\textbf{x})-\psi\delta^{aj}\sigma_{ij},
\eea
where $\dt$ denotes tensor perturbations.
Using the decomposition in \eqref{gauge-field-perti}, However, we find that the genuine tensor perturbation of the gauge field is invariant under this diffeomorphism 
\be
\tg_{ij}(t,\textbf{x})\mapsto \tg_{ij}(t,\textbf{x}).
\ee
Therefore the canonically normalized field corresponding to the $\tg_{ij}$ field (and not $\dt A^a_i$) is an independent degree of freedom which should be quantized\footnote{Note that our canonically normalized field is different from ref.s \cite{Adshead:2013nka, Adshead:2016omu,Adshead:2013qp}, in which $\dt A^a_i$ has been quantized.}. 

Recalling that $\gamma_{ij}$ appears only with derivatives in the quadratic perturbed Einstein-Hilbert action, we find that its contribution to the action and the field equation of $\gamma_{ij}$ are both invariant under the transformation \eqref{g-trans}. On the other hand, it may seem that $\gamma_{ij}$ can appear without any derivatives in the Yang-Mills terms, e.g. in $\bar{F}^a_{ij}\bar{F}^a_{kl}\gamma^{ik}\gamma^{jl}$ and $\bar{F}^a_{0i}\dt F^a_{0j}\gamma^{ij}$. That is not the case, however. As we will show in \ref{field-eq}, these terms are canceled by the metric induced part in the gauge field, $\psi\dt\!e^a_{~i}$. In other words, having a diffeomorphism and gauge invariant theory \eqref{action}, the gravitational wave $\gamma_{ij}$ can only appear with a derivative. Therefore, both the second order action and the field equation of $\gamma_{ij}$ are invariant under the diffeomorphism \eqref{g-trans}. In fact, the diffeomorphism invariance in our theory forbids the existence of an effective mass term for the graviton.

\subsection{Field equations}\label{field-eq}

Considering the symmetric, traceless and divergence-free fluctuations, the perturbed metric is 
\be
\dt g_{ij}=a^2e^{\gamma_{ij}} \quad \textmd{and} \quad \dt g_{0i}=\dt g_{00}=0,
\ee
where $\gamma_{ii}=\partial_j\gamma_{ij}=0$, while using \eqref{gauge-field-pert}, the perturbed gauge field is
\be\label{delta-T-A}
\dt A^a_{i}=a\delta^{aj} \bigg(\tg_{ij}+\frac{\psi}{2}[\gamma_{ij}+\frac12\gamma_{ik}\gamma_{kj}]\bigg) \quad \textmd{and} \quad \dt A^a_0=0.
\ee 
The perturbed energy-momentum tensor has the following tensor part
\be
\dt T_{ij}=\bar P\dt g_{ij}(t,\textbf{x})+a^2\pi^{T}_{ij}(t,\textbf{x}),
\ee
where $\pi^{T}_{ij}$ is the traceless transverse part of the anisotropic stress given as
\be\label{piT-}
\pi^T_{~ij}=H\psi\bigg((2\xi_{\psi}\xi-1)H\tg_{ij}-\dot{\tg}_{ij}
-\xi_{\psi}a^{-1}\epsilon^{ilk}\partial_l\tg_{jk}\bigg).
\ee
The non-vanishing $\pi^T_{~ij}$ modifies the field equation of $\gamma_{ij}$
\be \label{eq-gamma}
\ddot \gamma_{ij}+3H \dot \gamma_{ij}-a^{-2}\partial^2\gamma_{ij}=2\pi^T_{ij}\,.
\ee
The field equation of $\tg_{ij}$ is given by the second order action of the tensor modes. Here we have the final equation and the details are presented in \ref{app-action}
\be\label{eq-tg}
\ddot{\tg}_{ij}+3H\dot{\tg}_{ij}-a^{-2}\partial^2\tg_{ij}+2(1+\xi\xi_{\psi}-\frac12\epsilon)H^2\tg_{ij}-2(\xi+\xi_{\psi})Ha^{-1}\epsilon^{ilk}\partial_{l}\tg_{jk}\simeq0.
\ee

Here, we summarize the noteworthy features of the tensor field equations \eqref{eq-gamma} and \eqref{eq-tg}.
\begin{enumerate} 
\item{The gravitational wave, $\gamma_{ij}$, only appears with (at least) a derivative acting on it\footnote{Note that considering $\dt A^a_{~i}$ as the canonically normalized field (and not the gauge invariant $a\tg_{ij}$) leads to some effective mass terms for the graviton which breaks the diffeomorphism invariance (See e.g. \cite{Adshead:2013qp}).}. Therefore, similar to the inflationary models with no tensor sources, there is a shift symmetry for the $\gamma_{ij}$. The action's invariance under \eqref{large-x} is the direct consequence of diffeomorphism invariance of our theory.}
\item{The tensor sector has four dynamical degrees of freedom, two for each of $\gamma_{ij}$ and $\tg_{ij}$.}
\item{A constant $\gamma_{ij}$ and $\tg_{ij}=0$ (with $\pi^T_{ij}=0$) is a solution of the field equations.}
\item{Comparing to the usual (scalar) inflationary models the field equation of $\gamma_{ij}$ is modified by a
nonvanishing anisotropic inertia which is a function of $\tg_{ij}$.}
\item{In both $\pi^T_{ij}$ and field equation of $\tg_{ij}$ in \eqref{eq-tg} there are parity odd terms which take different signs for the two polarizations of the tensor modes,
leading to the existence of intrinsic chiral gravitons.}
\item{The spin-2 fluctuation, $\tg_{ij}$, has a positive mass equal to $2(1+\xi\xi_{\psi})H^2$. Thus, it is a massive field and exponentially damps after horizon crossing.}
\end{enumerate}


In the following, we present the solutions of the tensor perturbations, including the gravitational waves and the spin-2 perturbations of the gauge field. We first focus on the solution of the gauge field's spin-2 perturbation, $\tg_{ij}$. Next, we turn to the gravitational waves, $\gamma_{ij}$, and present its homogeneous part and inhomogeneous solution sourced by the gauge field.

\subsection{Spin-2 fluctuations of gauge field}

In Fourier space, the spin-2 fluctuation of the gauge field, $\tg_{ij}$, can be expanded in helicity basis as
\be\label{Fourier}
\tg_{ij}(\tau,\textbf{x})=\sum_{\lambda=\pm}\int \frac{d^3k}{(2\pi)^3}\bigg[ \tilde\gamma_{\lambda}(\tau,\textbf{k})\hat b(\textbf{k},\lambda)\textbf{e}_{ij}(\hat{k},\lambda)e^{i\textbf{k}.\textbf{x}}+\tilde\gamma^{*}_{\lambda}(\tau,\textbf{k})\hat b^{\dagger}(\textbf{k},\lambda)\textbf{e}^{*}_{ij}(\hat{k},{\lambda})e^{-i\textbf{k}.\textbf{x}}\bigg],~~~~~~~~~~
\ee
where $e_{ij}(\hat{k},\lambda)$ are the time-independent polarization tensors satisfying the conditions
\bse
\begin{align}
&\textbf{e}_{ij}(\hat k,\lambda)\textbf{e}^{*}_{~ij}(\hat k,\tilde{\lambda})=2\delta^{\lambda\tilde{\lambda}}, \\
&\epsilon^{ilk}\hat{k}_l\textbf{e}_{jk}(\hat k,\pm )=\pm i \textbf{e}_{ij}(\hat k,\pm),
\end{align}
\ese
in which $\lambda=\pm$ corresponding to the $\pm2$ helicity states.
Moreover, for a $\textbf{k}$ in the $z$-direction, the polarization tensors have components
\be
\textbf{e}_{11}(\hat{z},\pm)=-\textbf{e}_{22}(\hat{z},\pm)=\frac{1}{\sqrt{2}}, \quad \textbf{e}_{12}(\hat{z},\pm)=\textbf{e}_{21}(\hat{z},\pm)=\frac{\pm i}{\sqrt{2}} \quad \textmd{and} \quad \textbf{e}_{i3}(\hat{z},\pm)=\textbf{e}_{3i}(\hat{z},\pm)=0.
\ee
The creation and annihilation operators satisfy the standard commutation relations
\be\label{commutation}
[\hat b_{\textbf{k},\lambda},\hat b_{\textbf{k}',\tilde\lambda}^{\dag}]=(2\pi)^3\delta_{\lambda\tilde\lambda}\delta^{(3)}(\textbf{k}-\textbf{k}') \quad \textmd{and}  \quad [\hat b_{\textbf{k},\lambda},\hat b_{\textbf{k}',\tilde\lambda}]=[b^{\dag}_{\textbf{k},\lambda},b^{\dag}_{\textbf{k}',\tilde\lambda}]=0.
\ee
For later convenience, here we define $\mathrm{\th}_{\pm}(\tau,\textbf{k})$ as
\be
\mathrm{\th}_{\pm}(\tau,\textbf{k})\equiv \frac{\sqrt{2} a}{(2\pi)^\frac32}\tg_{\pm}(\tau,\textbf{k}),
\ee
which is the canonically normalized field corresponding to $\tg_{\pm}$. In terms of $\mathrm{\th}_{\pm}$ and re-definitions below
\be\label{re-def}
z=2ik\tau,\quad \kappa_{\pm}=\mp i\big(\xi+\xi_{\psi}\big) \quad \textmd{and} \quad \mu^2=\frac14-2\xi\xi_{\psi},
\ee
and after using the slow-roll relation $\mH\simeq-\frac{(1+\epsilon)}{\tau}$, we can write the field equation \eqref{eq-tg} as
\be\label{th-eqq}
\partial^2_{z}\mathrm{\th}_{\pm}(k,\tau)+(-\frac14+\frac{\kappa_{\pm}}{z}+\frac{1/4-\mu^2}{z^2})\mathrm{\th}_{\pm}(k,\tau)\simeq0.
\ee
The most general solutions of the above equation are Whittaker functions, $W_{\kappa,\mu}(z)$ and $M_{\kappa,\mu}(z)$. Recalling that in the asymptotic past, the $W_{\kappa,\mu}(2ik\tau)$ represents the positive frequency solution\footnote{For $\mid z\mid\rightarrow\infty$, the asymptotic from of the Whittaker functions are
\be
W_{\kappa,\mu}(z)\rightarrow z^{\kappa}e^{-z/2}, \quad M_{\kappa,\mu}(z)\rightarrow \Gamma(2\mu+1)\big(\frac{i(-1)^{\mu-\kappa}z^{\kappa}e^{-z/2}}{\Gamma({-\kappa+\mu+\frac12})}+\frac{z^{-\kappa}e^{z/2}}{\Gamma({-\kappa+\mu+\frac12})}\big) \quad \textmd{for} \quad \mid \arg z\mid<\frac32\pi.\nonumber
\ee
}
and imposing the usual Banch-Davis initial condition, we have
\be\label{tilde-h}
\mathrm{\th}_{\pm}(k,\tau)=\frac{e^{i\pi\kappa\!_{\pm}\!/2}}{\sqrt{2k}}~ W_{\kappa\!_{\pm},\mu}(2ik\tau),
\ee
in which we neglect a phase term. 
At large scales, $-k\tau\ll1$, we have $\mathrm{\th}_{\pm}(k,\tau)\propto \tau^{\frac12\pm \mu}$ where $\mu$ (which is given in \eqref{re-def}) is an imaginary quantity.\footnote{For $z\rightarrow0$, the asymptotic from of the $W_{\kappa,\mu}$ function is $W_{\kappa,\mu}(z)\rightarrow (1+i)\sqrt{z}\big(\frac{(2i)^{\mu}\Gamma(-2\mu)z^\mu}{\Gamma(\frac12-\kappa-\mu)}+\frac{(2i)^{-\mu}\Gamma(2\mu)z^{-\mu}}{\Gamma(\frac12-\kappa+\mu)}\big)$. } As a result, the spin-2 perturbation of the gauge field rapidly damps after horizon crossing. From the above, we can recognize three distinct regions in the evolution of $\mathrm{\th}_{\pm}(k,\tau)$:
\begin{enumerate}
\item{In the deep inside horizon limit, $-k\tau\gg1$, we have $\mathrm{\th}_{\pm}(k,\tau)\simeq\frac{1}{\sqrt{2k}}e^{-ik\tau}$.}
\item{For a short interwal before and at the visitinty of the horizon crossing, 
 the effective frequency of one of the polarization states in \eqref{th-eqq} becomes negative which leads to its \textit{temporary} tachyonic growth. }
\item{These modes have a positive mass term and at super horizon scales,$-k\tau\ll1$ , we have $\mathrm{\th}_{\lambda}(k,\tau)\propto a^{-\frac12\pm \mu}$. In fact, the spin-2 perturbations of the gauge field, $\tg_{\lambda}$, are heavy fields with a mass term given as $m^2_{\tg}/H^2=2(1+\xi\xi_{\psi})$. Therefore, its both polarization states decay with the same rate and oscillate,
scaling as $\tg_{\lambda}\propto a^{-\frac32\pm\sqrt{2\xi\xi_{\psi}-\frac14}}$.}
\end{enumerate}

\subsection{Graviton}\label{graviton-sub}

In this part, we compute the solution for the graviton which is now interacting with the spin-2 fluctuations of the gauge field. It is useful to decompose $\gamma_{ij}$ as 
\be\label{hG+hS}
\gamma_{ij}(\tau,\textbf{x})=\gamma^{G}_{ij}(\tau,\textbf{x})+\gamma^{S}_{ij}(\tau,\textbf{x}),
\ee
where $\gamma^{G}_{ij}$ and $\gamma^{S}_{ij}$ are the homogeneous and the particular parts of the solution respectively. 

\textit{i) Homogeneous solution:} In Fourier space, we can expand the graviton in the helicity basis as
\bse
\begin{align}
\gamma^G_{ij}(\tau,\textbf{x})=\sum_{\lambda=\pm}\int \frac{d^3k}{(2\pi)^3}\bigg[ \gamma^G_{\lambda}(\tau,\textbf{k})\hat a(\textbf{k},\lambda)\textbf{e}_{ij}(\hat{k},\lambda)e^{i\textbf{k}.\textbf{x}}+\gamma^{G*}_{\lambda}(\tau,\textbf{k})\hat a^{\dagger}(\textbf{k},\lambda)\textbf{e}^{*}_{ij}(\hat{k},{\lambda})e^{-i\textbf{k}.\textbf{x}}\bigg],\nonumber
\end{align}
\ese
where $\hat a^{\dagger}$  and $\hat a$ are the creation and annihilation operators of the graviton respectively.
It is useful to define $\mathrm{h}_{\pm}(\tau,\textbf{k})$ as
\be
\mathrm{h}^G_{\pm}(\tau,\textbf{k})\equiv \frac{1}{\sqrt{2}(2\pi)^{\frac32}}a\gamma^G_{\pm}(\tau,\textbf{k}),
\ee
which is the canonically normalized field corresponding to $\gamma_{\pm}$. Using \eqref{eq-gamma}, we find that the homogeneous part, $h^{G}_{_{R,L}}$, satisfies in
\be\label{eq-h---}
\mathrm{h}^{G''}_{\pm}+(k^2-(2-\epsilon)\mathcal{H}^2)\mathrm{h}^G_{\pm}= 0,
\ee
which is the same for both of the polarizations. The above equation indicates that $\mathrm{h}_{\pm}(\tau,\textbf{k})$ is \textit{unpolarized}, i.e. we have
\be
\mathrm{h}(\tau,\textbf{k})\equiv e^{\mp i\pi/4}\mathrm{h}^{G}_{\pm}(\tau,\textbf{k}),
\ee
 and has the field equation of a massless scalar field, similar to the standard scalar-inflation models. The initial condition of $\mathrm{h}(\tau,\textbf{k})$ is set by the Banch-Davis condition,\footnote{It is noteworthy to mention that the interaction terms between the graviton and the spin-2 field of the gauge field are negligible in the asymtotic past limit, $k\tau\rightarrow-\infty$.} \textit{i.e.} $\mathrm{h}(\tau,\textbf{k})\mathrm{h}^{*'}(\tau,\textbf{k})-\mathrm{h}'(\tau,\textbf{k})\mathrm{h}^*(\tau,\textbf{k})=i$. As a result, we find the wave function of the general solution as
\be\label{h-hankel}
\mathrm{h}(\tau,\textbf{k})\simeq \frac{\sqrt{-\pi\tau}}{2}e^{i(1+2\nu)\pi/4}H^{^{(1)}}_{\nu}(-k\tau),
\ee
where $\nu\simeq\frac32+\epsilon$ and $H^{^{(1)}}_{\nu}$ is the Hankel function of the first kind.
Using the asymptotic form of the Hankel function for $k\tau\rightarrow0, $\footnote{The asymptotic form of the Hankel function of the first kind for $k\tau\rightarrow0$ is given as $H^{^{(1)}}_{\nu}(-k\tau)\rightarrow\frac{-i}{\pi}\Gamma(\nu)\big(-\frac{k\tau}{2}\big)^{-\nu}$.} and $\Gamma(3/2)=\sqrt{\pi}/2$, we can find that $\mathrm{h}^G(\tau,k)\simeq ik^{-\nu}(-\tau)^{-\nu+\frac12}$. That indicates that after horizon crossing, the general solution of the gravitational wave, $\gamma^G(\tau,k)$, goes rapidly to a constant. In particular, on the large scales ($k\tau\ll1$) we have
\be\label{super-h}
e^{\mp i\pi/4}k^{\frac32}\gamma^G_{\pm}(\tau,k)\simeq (2\pi)^{\frac32} iH k^{-\epsilon},
\ee
in which we used the slow-roll relation $a\simeq(-\tau)^{-(1+\epsilon)}/H$. Using the relation 
\be\label{gammaij}
\gamma_{\pm}\equiv\frac{1}{\sqrt{2}}(\gamma_{11}\pm i\gamma_{12}),
\ee
 we can write $\gamma^G_{ij}$ as $\gamma^G_{11}(\tau,k)=\gamma^G_{12}(\tau,k)$, and $\gamma^G_{11}(\tau,k)=-\gamma^G_{22}(\tau,k)$.

\vskip 1cm

\textit{ii) Inhomogeneous solution:} The inhomogeneous solution for the gravitational waves is sourced by the spin-2 perturbations of the gauge field given in \eqref{tilde-h}. Hence it can be expanded in terms of $b_{\lambda}$ and $b^{\dag}_{\lambda}$ as
\bse
\begin{align}
\gamma^S_{ij}(\tau,\textbf{x})=\sum_{\lambda=\pm}\int \frac{d^3k}{(2\pi)^3}\bigg[ \gamma^S_{\lambda}(\tau,\textbf{k})\hat b(\textbf{k},\lambda)\textbf{e}_{ij}(\hat{k},\lambda)e^{i\textbf{k}.\textbf{x}}+\gamma^{S*}_{\lambda}(\tau,\textbf{k})\hat b^{\dagger}(\textbf{k},\lambda)\textbf{e}^{*}_{ij}(\hat{k},{\lambda})e^{-i\textbf{k}.\textbf{x}}\bigg],\nonumber
\end{align}
\ese
The inhomogeneous wave function in term of 
\be
\mathrm{h}^S_{\pm}(\tau,\textbf{k})\equiv \frac{1}{\sqrt{2}(2\pi)^{\frac32}}a\gamma^S_{\pm}(\tau,\textbf{k}),
\ee
 can be written as 
\be\label{h-sigma}
\mathrm{h}^{S}_{\pm}(\tau,k)=\int^{\tau}_{-\infty}G(\tau,\tau')S_{\pm}(\tau',k)d\tau',
\ee
where $G(\tau,\tau')$ is the retarded Green's function of Eqn. \eqref{eq-h---}, which in terms of $\mathrm{h}(\tau,\textbf{k})$ in \eqref{h-hankel} can be written as
\bea
\label{Green}
G(\tau,\tau')
=\frac{\mathrm{h}(\tau,\textbf{k})\mathrm{h}^*(\tau',\textbf{k})-\mathrm{h}(\tau',\textbf{k})\mathrm{h}^*(\tau,\textbf{k})}{\textsf{W}\big(\mathrm{h},\mathrm{h}^{*}\big)}\Theta(\tau'-\tau),
\eea
where $\textsf{W}(\mathrm{h}, \mathrm{h}^*)$ is the Wronskian of $\mathrm{h}$ and $\mathrm{h}^*$,
$\textsf{W}(\mathrm{h}, \mathrm{h}^*)=i$, while $\Theta(\tau-\tau')$ is the Heaviside step function. Moreover, inserting \eqref{tilde-h} in \eqref{piT-}, we have the explicit form of the source as
\be
S_{_\pm}(k,\tau)\simeq e^{\pm i\pi/4}\bigg(\frac{\bar{\rho}_{_{\rm YM}}}{\bar{\rho}}\bigg)^{\!\frac12}\frac{ e^{\pm(\xi+\xi_{\psi})\frac{\pi}{2}}}{\sqrt{k( 1+\xi^2_{\psi})}}\mH\bigg(-\partial_{\tau}+\xi_{\psi}(2\xi\mH\mp k)\bigg)W_{\kappa_{\pm},\mu}(2ik\tau),
\ee
which is weighted by the ratio of the energy of the gauge field to the total energy as $(\frac{\bar{\rho}_{_{\rm YM}}}{\bar{\rho}})^{\!\frac12}$.
It is interesting to notice that $S_{\pm}$ is proportional to $e^{\pm(\xi+\xi_{\psi})\frac{\pi}{2}}$ which for $\xi\sim\xi_{\psi}$, after recalling that during the slow-roll inflation $(\frac{\bar{\rho}_{_{\rm YM}}}{\bar{\rho}})^{\!\frac12}\sim \epsilon$, implies that $S_-\lesssim\mathcal{O}(\epsilon)$ while $S_+\sim 1$. Therefore, we find that $\mathrm{h}^{S}_{-}(\tau,k)$ is negligible while $\mathrm{h}^{S}_{+}(\tau,k)$ can be of the same order as $\mathrm{h}(\tau,k)$.

In order to find the super-horizon form of $\mathrm{h}^{S}_{+}(\tau,k)$, one needs to compute the integral \eqref{h-sigma} in $-k\tau\ll1$ limit.
As implies by \eqref{super-h}, at very large scales $\mathrm{h}(\tau,\textbf{k})$ is a pure imaginary quantity, i.e. $\mathrm{h}=-\mathrm{h}^*$. Therefore, we can write the Green's function as \footnote{Notice that we use the exact de-Sitter solution, $\mathrm{h}_{\rm deS}(\tau,\textbf{k})=\frac{1}{\sqrt{2k}}(1-\frac{i}{k\tau})e^{-ik\tau}$, for $\mathrm{h}(\tau',\textbf{k})$ which is inside the integral. In fact, $\mathrm{h}(\tau',\textbf{k})\simeq\mathrm{h}_{\rm deS}(\tau',\textbf{k})$ is a very good approximation inside the horizon and the integrand vanishes at $\rvert \tau'\rvert\ll1$ (has the asymptotic form like $\sqrt{-\tau'}$). Thus, that is a very good approximation for the Green's function in the regime of our interest.}  
\bea
\label{Green}
G(\tau,\tau')
=\frac{ 1}{\sqrt{2k}}\mathrm{h}(\tau,k)\bigg(\cos(k\tau')-\frac{\sin(k\tau')}{k\tau'}\bigg)\Theta(\tau'-\tau) \quad (\rvert k\tau\rvert\ll1).
\eea
As we see from the above, in the limit that $\tau'$ is very small too, we have $G(\tau,\tau')\simeq -\frac{i}{3\sqrt{2k}}\mathrm{h}(\tau,k)k^2\tau'^2$ and $S_{+}(\tau',k)\sim \frac{1}{\tau'^{2}}\times \sqrt{-\tau'}$. As a result, the integrand in \eqref{h-sigma} is proportional to $\tau'^{\frac12}$ in the limit that $\tau'\sim\tau\ll1$ which implies that the integral is only given by its value at the initial time, $|\tau'|=|\tau_{\rm in}|\gg1$. Thus, we can simply write $\mathrm{h}_{+}(\tau,k)$ in terms of the wave function $\mathrm{h}(\tau',k)$ and a time-independent coefficient as
\be
\mathrm{h}^S_+(\tau,k)\simeq \bigg(\frac{\bar{\rho}_{_{\rm YM}}}{\bar{\rho}}\bigg)^{\!\frac12}e^{ i\pi/4}\mathcal{G}(\xi,\xi_{\psi})\mathrm{h}(\tau,k),
\ee
where for $\xi\sim\xi_{\psi}$ which we are interested in ($\sqrt{2}<\xi_{\psi}<3$), the coefficient $\mathcal{G}(\xi,\xi_{\psi})$ is such that $(\frac{\bar{\rho}_{_{\rm YM}}}{\bar{\rho}})^{\!\frac12}\mathcal{G}(\xi,\xi_{\psi})\sim1$ \cite{Maleknejad:2016qjz}. \footnote{The time-independent coefficient $\mathcal{G}(\xi,\xi_{\psi})$ is given as 
\be
\mathcal{G}(\xi,\xi_{\psi})=\frac{ie^{(\xi+\xi_{\psi})\pi/2}}{k\sqrt{2(1+\xi_{\psi}^2)}}\int\mH(\tau')\bigg(\cos(k\tau')-\frac{\sin(k\tau')}{k\tau'}\bigg)\bigg(-\partial_{\tau'}+\xi_{\psi}(2\xi\mH(\tau')- k)\bigg)W_{\kappa_{+},\mu}(2ik\tau')d\tau'\bigg|_{\tau'=\tau_{\rm in}},\nonumber
\ee
where $\tau_{\rm in}$ is the initial conformal time. Recalling that at $\tau\rightarrow-\infty$, the $W$ function is $ W_{\kappa_{+},\mu}(2ik\tau)\propto e^{-ik\tau}$, we find that $\mathcal{G}(\xi,\xi_{\psi})$ is $k$-independent too.}
Note that the above equation is only a relation between wave functions, while their operators are uncorrelated. 
Finally, we have the super-horizon form for the gravitational waves as
\be\label{super-hs}
e^{-i\pi/4}k^{\frac32}\gamma^{S}_{_{+}}(\tau,k)\simeq (2\pi)^{\frac32} iH k^{-\epsilon}\bigg(\frac{\bar{\rho}_{_{\rm YM}}}{\bar{\rho}}\bigg)^{\!\frac12}\mathcal{G}(\xi,\xi_{\psi}) \quad  \textmd{and} \quad \gamma^{S}_{_{-}}(\tau,k)\simeq  0,
\ee
which is a constant and is only non-zero for the state with helicity $+2$.  Using the relation \eqref{gammaij},
we find $\gamma^S_{ij}$ as $\gamma^S_{11}(\tau,k)=e^{i\pi/2}\gamma^S_{12}(\tau,k)$, and $\gamma^S_{11}(\tau,k)=-\gamma^S_{22}(\tau,k)$.

Finally, from the combination of \eqref{super-h} and \eqref{super-hs}, we find the two-point function of $\gamma_{\lambda}$ at late times as
\be\label{gamma-power}
\langle \gamma_{\lambda}(\textbf{q})\gamma_{\lambda'}(\tilde{\textbf{q}})\rangle=(2\pi)^3\delta^{(3)}(\textbf{q}+\tilde{\textbf{q}})P^{\lambda}_{\gamma}(q)\delta^{\lambda\lambda'},
\ee
where using \eqref{P-vac} and the fact that $\gamma^G$ and $\gamma^S$ are uncorrelated with each other, we have
\be\label{p+p-}
P_{\gamma}^{+}(q)\simeq \bigg(1+\frac{\bar{\rho}_{_{\rm YM}}}{\bar{\rho}}\mathcal{G}^2(\xi,\xi_{\psi})\bigg)P_{\gamma}^{\rm vac}(q) \quad \textmd{and} \quad P_{\gamma}^{-}(q)\simeq P_{\gamma}^{\rm vac}(q).
\ee
As we see, $P^{+}_{\gamma}$ and $P^{-}_{\gamma}$ are not equal and we can define the dimensionless chirality factor as below 
\be
X\equiv\frac{P^{+}_{\gamma}(q)-P^{-}_{\gamma}(q)}{P^{\rm vac}_{\gamma}(q)}\simeq \frac{\bar{\rho}_{_{\rm YM}}}{\bar{\rho}}\mathcal{G}^2(\xi,\xi_{\psi}),
\ee
which indicates that the chirality factor is proportional to $\frac{\bar{\rho}_{_{\rm YM}}}{\bar{\rho}}e^{(\xi+\xi_{\psi})\pi/2}$.

\section{Soft gravitions and tensor consistency relations}\label{soft}

In this section, we focus on the super-horizon behavior of the tensor perturbations in the axion-gauge field setup and discuss the adiabatic nature of these modes. Finally, we turn to the tensor consistency relations and investigate the $n$-point functions in the presence of a soft graviton with a small momentum. 

\subsection{Adiabatic modes}

Up to now, we find that both polarization states of the spin-2 perturbations of the gauge field, $\tg_{ij}$, damp like $a^{-3/2}$ shortly after the horizon crossing and thus vanish in the limit of $k/\mH\ll1$. Therefore the anisotropic stress also damps, i.e. $\pi^{T}_{ij}\propto a^{-3/2}$. However, due to the temporary tachyonic growth of its $+2$-helicity state, it acts like an impulse function for the graviton and sources the $+2$-helicity state of the gravity wave. As a result, one can decompose the gravity wave into two incoherent parts, $\gamma^G_{ij}$ and $\gamma^S_{ij}$ which are the homogeneous and the inhomogeneous (sourced by the gauge-field) solutions of the gravity waves. As we found in section \ref{graviton-sub}, both of $\gamma^G_{ij}$ and $\gamma^S_{ij}$ are constant for $k/\mathcal{H}\ll1$. In the following, we show that both of these solutions are adiabatic, as we expect from the damping behavior of $\pi^T_{ij}$ at super-horizon scales. 

\textit{i) Homogeneous solution:}

The homogeneous solution of the graviton is unpolarized which in Fourier space and for $q/\mH\ll1$ limit is a constant given in \eqref{super-h}. This long-wavelength mode has the following form in the real space
\be\label{gamma-hom}
\gamma^G_{ ij}(t,\textbf{x})=c\left(
\begin{tabular}{ccc}
$\cos(qz)$ & $\cos(qz)$ & 0 \\
$\cos(qz)$ & -$\cos(qz)$ & 0  \\
0 & 0 & 0\\
\end{tabular}
\right),
\ee
where $c$ is a constant and we have taken the z-axis in the direction of $\textbf{q}$.
In the limit that $q\rightarrow0$, we can neglect the gradients and arrive at
\be\label{gamma-hom}
\gamma^G_{ ij}(t,\textbf{x})\simeq2\sigma_{ij} \quad \textmd{where} \quad \sigma_{ij}=\frac{c}{2}\left(
\begin{tabular}{ccc}
$1$ & $1$ & 0 \\
$1$ & -$1$ & 0  \\
0 & 0 & 0\\
\end{tabular}
\right).
\ee

\textit{ii) inhomogeneous solution:}

In Fourier space and for $q/\mH\ll1$ limit, the polarized inhomogeneous solution of the graviton is constant (eq. \eqref{super-hs})
\be\label{super-gammas}
\gamma^{S}_{_{+}}(\tau,k)= i \sigma_{+} \quad  \textmd{and} \quad \gamma^{S}_{_{-}}(\tau,k)\simeq  0,
\ee
where $\sigma_+$ is a constant in $\mathbb{R}$. The real space form of $\gamma^S_{ij}$ is
\be\label{gamma-in-hom}
\gamma^S_{ ij}(t,\textbf{x})=\tilde c\left(
\begin{tabular}{ccc}
$\cos(qz)$ & $\sin(qz)$ & 0 \\
$\sin(qz)$ & -$\cos(qz)$ & 0  \\
0 & 0 & 0\\
\end{tabular}
\right),
\ee
where $\tilde c$ is a constant and again the z-axis is taken in the direction of \textbf{q}. It is interesting to notice that the above $\gamma^S_{ij}$ is a circularly polarized standing gravitational wave which adding it to the FRW metric gives
\be
ds^2=-dt^2+a^2(t)\bigg(\delta _{ij}+\gamma_{ij}^S(t,\textbf{x})\bigg)dx^idx^j,
\ee
which is a Bianchi type VII$_0$ metric (see \eqref{Bianchi-VII0}). Thus the inhomogeneous solution of the graviton is a \textit{circularly polarized} adiabatic tensor perturbation. 

In the limit that $q\rightarrow0$ and after neglecting the gradients, we have
\be\label{gamma-hom}
\gamma^S_{ ij}(t,\textbf{x})\simeq2\tilde{\sigma}_{ij} \quad \textmd{where} \quad \tilde\sigma_{ij}=\frac{\tilde c}{2}\left(
\begin{tabular}{ccc}
$1$ & $0$ & 0 \\
$0$ & -$1$ & 0  \\
0 & 0 & 0\\
\end{tabular}
\right).
\ee

To summarize, we showed that the gravitational wave in the presence of the SU(2) gauge field is an adiabatic mode, given as 
$$\gamma_{ij}=\gamma_{ij}^G+\gamma_{ij}^S.$$ 
The homogeneous solution is exactly the solution of tensor perturbations of the standard scalar inflationary models. The inhomogeneous part which is sourced by the gauge field is also a constant at super-horizon scales. However, it is polarized and incoherent with the homogeneous solution. 
The total long-wavelength mode, $\gamma_{ij}(\textbf{q})$, is equivalent to the following coordinate transformation for the shorter modes, $\zeta_{\textbf{k}}$ and $\gamma_{ij}(\textbf{k})$, which crosses the horizon much later than it ($q/k\rightarrow0$)
\be\label{change-coor}
x^i \rightarrow \tilde x^i=x^i+\delta x^i \quad \textmd{where} \quad \delta x^i=\frac12\gamma_{ij}x^j.
\ee
Here we emphasize that our tensor perturbations are the same as the tensor sector in the gauge-flation \cite{Maleknejad:2011jw, Maleknejad:2011sq} and chromo-natural inflation \cite{Adshead:2012kp, Adshead:2013qp}. \footnote{As it has been shown explicitly in ref. \cite{Maleknejad:2012fw}, the linear tensor and vector perturbations are exactly the same in the gauge-flation and chromo-natural inflation models.} Hence, the gravity waves in these models are also described by the above adiabatic modes. \footnote{Notice that our result is different from the statement of ref. \cite{Bordin:2016ruc} on the nature of gravity waves in these models.} It is shown in \cite{Maleknejad:2011jr, Maleknejad:2013npa} that these models satisfy the cosmic no-hair conjecture and the initial anisotropies damp out exponentially in a few e-folds. That is in agreement with the adiabatic nature of the gravity waves and exponential decay of the tensor anisotropic stress at very large scales which are necessary for the cosmic no-hair conjecture to hold.

\subsection{Consistency relations for gravitational waves}

The adiabatic nature of the gravity waves implies that the change of the short distance n-point correlation function in the presence of $\gamma$ is
\be
\delta_{\gamma}\langle \zeta(\tilde{x}_1)\zeta(\tilde{x}_2)\cdots\zeta(\tilde{x}_n) \rangle=\sum_{\rm I=1}^{n}\delta \vec{x}_{\rm I}.\vec{\nabla}_{\rm I}\langle \zeta(x_1)\zeta(x_2)\cdots\zeta(x_n) \rangle +\cdots,
\ee
which neglecting the gradients of the long-wavelength mode gives 
\be
\delta_{\gamma}\langle \zeta(\tilde{x}_1)\zeta(\tilde{x}_2)\cdots\zeta(\tilde{x}_n) \rangle_{\textbf{q}\rightarrow0}\simeq-\frac12 \prod^n_{\rm I=1}\int e^{i\textbf{k}_{\rm I}.\textbf{x}_{\rm I}}\frac{d^3k_{\rm I}}{(2\pi)^3}\sum^n_{{\rm J}=1} \gamma_{ij}(\textbf{q}) k_{{\rm J}i}\partial_{k_{{\rm J}j}}\langle \zeta_{\textbf{k}_1}\zeta_{\textbf{k}_2}\cdots\zeta_{\textbf{k}_n} \rangle.
\ee
Therefore, the (n+1)-point correlation function including the long-wavelength mode is 
\bea
&&\langle \gamma_{ij}(x) \zeta(\tilde{x}_1)\zeta(\tilde{x}_2)\cdots\zeta(\tilde{x}_n) \rangle_{\textbf{q}\rightarrow0}\simeq \nonumber\\
&&-\frac12 \int  e^{i\textbf{q}.\textbf{x}} \frac{d^3q}{(2\pi)^3}\prod^n_{\rm I=1}\int e^{i\textbf{k}_{\rm I}.\textbf{x}_{\rm I}}\frac{d^3k_{\rm I}}{(2\pi)^3} \sum^n_{{\rm J}=1} \bigg\langle \gamma_{ij}(-\textbf{q})\gamma_{mn}(\textbf{q}) k_{{\rm J}m}\partial_{k_{{\rm J}n}}\langle \zeta_{\textbf{k}_1}\zeta_{\textbf{k}_2}\cdots\zeta_{\textbf{k}_n} \rangle\bigg\rangle,\nonumber
\eea
where using \eqref{gamma-power}, we can write the expectation value as
\bea
&&\sum_{\lambda,\lambda'=\pm}\textbf{e}_{ij}(-\hat{q},\lambda)\textbf{e}_{mn}(\hat{q},\lambda')\bigg\langle \gamma_{\lambda}(-\textbf{q})\gamma_{\lambda'}(\textbf{q}) k_{{\rm J}m}\partial_{k_{{\rm J}n}}\langle \zeta_{\textbf{k}_1}\zeta_{\textbf{k}_2}\cdots\zeta_{\textbf{k}_n} \rangle\bigg\rangle\nonumber\\
&&=(2\pi)^{3}\sum_{\lambda=\pm}\textbf{e}_{ij}(-\hat{q},\lambda) \delta^{(3)}(0)P_{\gamma}^{\lambda}(q) \textbf{e}_{mn}(\hat{q},\lambda) k_{{\rm J}m}\partial_{k_{{\rm J}n}}\langle \zeta_{\textbf{k}_1}\zeta_{\textbf{k}_2}\cdots\zeta_{\textbf{k}_n} \rangle.
\eea
Finally, we find 
\be\label{Pol-Malda}
\langle \gamma_{\lambda}(\textbf{q}) \zeta_{\textbf{k}_1}\zeta_{\textbf{k}_2}\cdots\zeta_{\textbf{k}_n} \rangle'\simeq-P^{\lambda}_{\gamma}(q)\sum_{\rm I=1}^n \epsilon_{ij}(\hat{q},\lambda) k_{{\rm I}i}\partial_{k_{{\rm I}j}}\langle \zeta_{\textbf{k}_1}\zeta_{\textbf{k}_2}\cdots\zeta_{\textbf{k}_n} \rangle' \quad \textmd{for} \quad q\rightarrow0 ,
\ee
which is different for each polarization state of the gravitational wave, i.e.
\bea
\langle \gamma_{+}(\textbf{q}) \zeta_{\textbf{k}_1}\zeta_{\textbf{k}_2}\cdots\zeta_{\textbf{k}_n} \rangle'_{q\rightarrow0}&\simeq&-\bigg(1+\frac{\bar{\rho}_{_{\rm YM}}}{\bar{\rho}}\mathcal{G}^2(\xi,\xi_{\psi})\bigg)P^{\rm vac}_{\gamma}(q)\sum_{\rm I=1}^n \epsilon_{ij}(\hat{q},+) k_{{\rm I}i}\partial_{k_{{\rm I}j}}\langle \zeta_{\textbf{k}_1}\zeta_{\textbf{k}_2}\cdots\zeta_{\textbf{k}_n} \rangle' ,\nonumber\\
\langle \gamma_{-}(\textbf{q}) \zeta_{\textbf{k}_1}\zeta_{\textbf{k}_2}\cdots\zeta_{\textbf{k}_n} \rangle'_{q\rightarrow0}&\simeq&-P^{\rm vac}_{\gamma}(q)\sum_{\rm I=1}^n \epsilon_{ij}(\hat{q},-) k_{{\rm I}i}\partial_{k_{{\rm I}j}}\langle \zeta_{\textbf{k}_1}\zeta_{\textbf{k}_2}\cdots\zeta_{\textbf{k}_n} \rangle' .\nonumber
\eea
To summarize, due to the adiabatic nature of long-wavelength gravitational waves, a system of short modes with and without the long mode are related by an anisotropic change of coordinate of the form \eqref{change-coor}. Since the long-wavelength mode in the presence of the SU(2) gauge field is polarized, such inflationary models satisfy in a polarized version of Maldacena's consistency relation as \eqref{Pol-Malda}. In fact, we have
\be
\frac{\langle \gamma_{+}(\textbf{q}) \zeta_{\textbf{k}_1}\zeta_{\textbf{k}_2}\cdots\zeta_{\textbf{k}_n} \rangle'-\langle \gamma_{-}(\textbf{q}) \zeta_{\textbf{k}_1}\zeta_{\textbf{k}_2}\cdots\zeta_{\textbf{k}_n} \rangle'}{\langle \gamma_{-}(\textbf{q}) \zeta_{\textbf{k}_1}\zeta_{\textbf{k}_2}\cdots\zeta_{\textbf{k}_n} \rangle'}|_{q\rightarrow0}\sim \frac{\bar{\rho}_{_{\rm YM}}}{\bar{\rho}}\mathcal{G}^2(\xi,\xi_{\psi}),
\ee
which is of the same order of the chirality factor in the power spectrum. In fact, regardless of the value of $n$, the dimensionless chirality factor above is of the order of $\frac{\bar{\rho}_{_{\rm YM}}}{\bar{\rho}}\mathcal{G}^2(\xi,\xi_{\psi})$. The chiral n-point functions provides a robust observational feature for the presence of the primordial gauge fields.

\section{ Conclusion} \label{conclude}

In this work, we studied the long wavelength tensor modes, adiabatic nature of the gravity waves and tensor consistency relation in the axion-gauge field inflationary models in which the graviton, $\gamma_{ij}$, is coupled to the spin-2 perturbation of the SU(2) gauge field, $\tilde\gamma_{ij}$. The tensor perturbation of the gauge field produces an anisotropic inertia for the graviton which damps like $a^{-3/2}$ shortly after the horizon crossing. Thus, as one may expect from the Weinberg's theorem, and we showed explicitly here, the gravity waves are adiabatic and become constant for $k\tau\ll1$. Its homogeneous part of the solution is equal to the prediction of the standard scalar inflationary models for $\gamma_{ij}$. On the other hand, the inhomogeneous part of the gravity wave is polarized, incoherent from the homogeneous part and proportional to $\frac{\bar{\rho}_{_{\rm YM}}}{\bar{\rho}}$. Therefore, the total gravity wave is partially polarized.
It is interesting to mention that for $k\tau\ll1$, the inhomogeneous solution of $\gamma_{ij}$, which is time-independent, can be described as a Bianchi type VII$_0$ space-time, at all gradient expansion order.  Since the tensor perturbations in our setup, gauge-flation, and chromo-natural inflation are the same, our arguments are correct for those models as well. Besides, we expect that our general results are correct for any covariant axion-gauge field setup as far as the spin-2 fluctuation of the gauge field is massive, i.e.  $m^2_{\tilde \gamma}>\frac{9H^2}{4}$.

Due to the adiabatic nature of long-wavelength gravity waves in the axion-gauge field setup, a system of short modes with and without the long mode are related by an anisotropic change of spatial coordinate. Since the long-wavelength mode in the presence of the SU(2) gauge field is polarized, such inflationary models satisfy in a polarized version of Maldacena's consistency relation. In particular, defining a chirality factor given as $\frac{\langle \gamma_{+}(\textbf{q}) \zeta_{\textbf{k}_1}\zeta_{\textbf{k}_2}\cdots\zeta_{\textbf{k}_n} \rangle-\langle \gamma_{-}(\textbf{q}) \zeta_{\textbf{k}_1}\zeta_{\textbf{k}_2}\cdots\zeta_{\textbf{k}_n} \rangle}{\langle \gamma_{-}(\textbf{q}) \zeta_{\textbf{k}_1}\zeta_{\textbf{k}_2}\cdots\zeta_{\textbf{k}_n} \rangle}|_{q\rightarrow0}$, we find it proportional to $\frac{\bar{\rho}_{_{\rm YM}}}{\bar{\rho}}\mathcal{G}^2(\xi,\xi_{\psi})$. In fact, the parity violation of the (n+1)-point functions including the soft graviton is of the same order as the chirality factor of the power spectrum of $\gamma_{ij}$. For typical values that used in \cite{Maleknejad:2016qjz}, i.e. $\xi, \xi_{\psi}\sim1$ and $\frac{\bar{\rho}_{_{\rm YM}}}{\bar{\rho}}\lesssim\epsilon^2$, this quantity can be of the order of $0.1-1$. We emphasize that the scalar perturbations in the presence of the SU(2) gauge fields are more complicated. In particular, the curvature perturbation is not adiabatic, and have a slow-roll suppressed entropy perturbation \cite{Maleknejad:2016qjz, Adshead:2016omu}. Therefore, the scalar $(n+1)$-point functions in these setups violate Maldacena's consistency relation which we leave for future work. The chiral $(n+1)$-point functions with a soft graviton provide a robust observational feature for the contribution of the primordial non-Abelian gauge fields in the physics of the early universe, which can be tested by the future ambitious CMB and LSS observations.     

\section*{\small Acknowledgment}

It is a pleasure to thank Mohammad M. Sheikh-Jabbari and Amjad Ashoorioon for their comments on the draft. The author would also like to thank Lorenzo Bordin and Paolo Creminelli for discussions on the topic analyzed in this work.

\appendix
\section{Gauge-invariant Combinations}\label{gauge-invariant} 

In this appendix, for self-completeness, we present the gauge-invariant combinations which can be constructed from an SU(2) gauge field.  Here with a slightly different parametrization, we essentially follow the approach presented in \cite{Maleknejad:2016qjz}. In section \ref{sec-gauge-invariant}, we showed that the first order $\delta A^a_{~\mu}(t,\textbf{x})$ can be decomposed into a space-time induced term, $\delta_{\rm x} A^a_{~\mu}=\psi \delta e^a_{~\mu}$, and the genuine gauge field fluctuations with 12 components, $\delta_{\rm g} A^a_{~\mu}$, as
\be
\delta A^a_{~\mu}(t,\textbf{x})=\delta_{\rm g} A^a_{~\mu}(t,\textbf{x})+\psi \delta e^a_{~\mu}(t,\textbf{x}).
\ee
Thus, $\delta A^a_{~\mu}(t,\textbf{x})$ can be decomposed as
\bse \label{gauge-field-pert-Appn}
\begin{align}
\delta A^a_{~i}&=a\delta^a_i \delta\psi+\delta^{aj}\big(\partial_{ij}\tilde{Z}+a\partial_i v_j+a\tg_{ij}\big)+a\psi\epsilon^{a~j}_{~i}\big(\g\partial_{j}(\tilde{Z}-Z)+aw_j\big)+\psi \delta e^a_{~i},\\
\delta A^a_{~0}&=\delta^{k}_a\partial_kY+a\delta_a^j u_j+\psi \delta e^a_{~0},
\end{align}
\ese
in which $\{\dd, Y, \tilde Z, Z\}$ parametrize scalar perturbations, $\{u_i, v_i, w_i\}$ are vector perturbations
$$\partial_iu_i=\partial_iv_i=\partial_iw_i=0,$$
and $\tg_{ij}$ is the SU(2) gauge field's is the symmetric, traceless and divergence-free tensor fluctuation
$$\tg_{ij}=\partial_i \tg_{ij}=0.$$
The most general perturbed FRW metric can be parametrized as below
\bea\label{metric-pert-app}%
&&ds^2=-(1+2A)dt^2+2a(\partial_iB+V_i)dx^idt+a^2\bigg((1-2C)\delta_{ij}+2\partial_{ij}E+2\partial_{(i}W_{j)}\nonumber\\
&&+\gamma_{ij}+\frac12\gamma_{ik}\gamma_{jk}\bigg)dx^idx^j\,,
\eea
where $A,\ B,\ C$ and $E$ are scalar fluctuations, $V_i,\ W_i$ are vector perturbations and finally $\gamma_{ij}$ is the tensor fluctuation of the metric. For this perturbed metric, we can choose the first order perturbed tetrad, $\delta e^a_{\mu}$, as
\bse \label{tetrad-pert-Appn}
\begin{align}
\delta e^a_{~i}&=a\bigg(-C\delta^{a}_{i}+\delta^{aj}(\partial_{ij}E+\partial_{(i}W_{j)}+\frac12[\gamma_{ij}+\frac12\gamma_{ik}\gamma_{jk}]\bigg),\\
\delta e^a_{~0}&=\delta^{aj}(a\partial_j\dot{E}+V_j).
\end{align}
\ese
Moreover, the linear order perturbed energy-momentum tensor around a background perfect fluid can be decomposed as %
\bse
\begin{align}
\delta T_{ij}=&\bar P\delta g_{ij}+a^2\left(\delta_{ij}(\delta P-\frac13\nabla^2\pi^S)+\partial_{ij}\pi^S
+2\partial_{(i}\pi^V_{j)}+\pi^T_{ij}\right)\,,\\
\delta T_{i0}=&\bar P\delta g_{i0}-(\bar \rho+\bar P)(\partial_i\delta u+\delta u_i^V)\,,\\
\delta T_{00}=&-\bar\rho\delta g_{00}+\delta \rho\,,
\end{align}
\ese
where $\bar\rho$ and $\bar P$ are the background energy and pressure densities. Departures from the perfect fluid form of the
energy-momentum tensor are characterized in terms of $\pi^S$, $\pi^V_i$, $\pi^T_{ij}$ which represent the \textit{anisotropic inertia}, as well as $\delta u_i^V$ which is the vorticity. They satisfy the following conditions
$$\partial_i\pi^V_i=\partial_i\pi^T_{ij}=\partial_i\delta u_i^V=0.$$

 Now we are ready to construct the gauge invariant combinations of each sector. Since $\{\dd, Y, \tilde Z, Z, u_i, v_i, w_i, \tg_{ij}\}$ are by definition invariant under the space-time gauge transformations, we need only to construct combinations which are invariant under the internal gauge field transformations
\be
\delta A^a_{\mu}\mapsto  \delta A^a_{\mu}+\frac{1}{\g}\partial_\mu\lambda^a+\epsilon^a_{~bc}A^b_{\mu}\lambda^c.
\ee 

\vskip 0.5cm
Scalar modes: In the scalar sector of the perturbations, we have
$\dd$, $Y$, $Z$ and $\tilde{Z}$ from the perturbations of the gauge field.
Under the action of the internal gauge field transformation of the form \eqref{gua}, the gauge field perturbations transform as
\be\label{A-g}
\begin{split}
\dd \mapsto \dd\,& ,\qquad
Y \mapsto Y+\frac{1}{\g}\dot{\lambda}\,,\\
Z\mapsto Z\,&,\qquad
\tilde{Z}\rightarrow \tilde{Z}+\frac{1}{\g}\lambda\,.
\end{split}
\ee
Therefore, the scalar gauge invariant combinations of the gauge field are
\be
\dd=\dd,\quad  M=\frac{\g^2\phi^3}{a^2}Z \quad \textmd{and} \quad \tilde{M}=\dot\phi(\dot{\tilde{Z}}-Y)\,.
\ee

\vskip 0.5cm
Vector modes: In the vector sector, we have $u_i$, $v_i$ and $w_i$ which under the infinitesimal gauge transformation \eqref{gua}, they change as
\be
u_i \mapsto u_i+\frac{1}{a\g}\dot{\lambda}_V^i\,,\quad
v_i \mapsto v_i+\frac{1}{a\g}\lambda_V^i\,,\quad
w_i \mapsto w_i+\frac1a\lambda_V^i\,.
\ee
Hence, we can construct two gauge invariant vector perturbations
\bea
\mathcal{U}_i=u_i-\frac{1}{a\g}{(aw_i\dot{)}}\,\quad \textmd{and} \quad \mathcal{V}_i=v_i-\frac1\g w_i\,.
\eea
Moreover, the gauge-invariant combination corresponding to the metric perturbations
\be
\mathcal{Z}_i=\dot{W}_i-\frac1aV_i.
\ee

\vskip 0.5cm
Tensor modes: In the tensor sector we have two gauge invariant spin-2 perturbations, $\gamma_{ij}$ and $\tg_{ij}$. 
The tensor fluctuation of the metric (gravity wave) is parametrized by $\gamma_{ij}$ and the spin-2 perturbation of the gauge field is presented by $\tg_{ij}$.

\section{Quadratic action of tensor modes}\label{app-action}

In this appendix, we work out the quadratic action of the symmetric traceless tensor perturbations.
Focusing on the tensor perturbations, we have the perturbed metric
\be
g_{00}=-1, \quad g_{0i}=0 \quad \textmd{and} \quad g_{ij}=a^2e^{\gamma_{ij}},
\ee
where $\gamma_{ij}(t,\vec{x})$ is a symmetric traceless tensor 
$$\gamma_{ii}(t,\vec{x})=\partial_{j}\gamma_{ij}(t,\vec{x})=0,$$
and the inverse metric is
\be\label{inverse-metric}
g^{00}=-1, \quad g^{0i}=0 \quad \textmd{and} \quad g^{ij}=a^{-2}e^{-\gamma_{ij}}.
\ee
The perturbed gauge field has the tensor fluctuations of the form 
\be\label{A-tensor}
A^a_{~i}=0 \quad \textmd{and} \quad  A^a_{~i}=a\psi \delta^{aj}e^{X_{ij}+\frac12\gamma_{ij}},
\ee
in which we used $e^a_{i}=\delta^{aj}e^{\frac12\gamma_{ij}}$ and $X_{ij}(t,\vec{x})$ is related to the genuine tensor modes of the gauge field as
\be\label{Xtog}
\tg_{ij}(t,\vec{x})\equiv \psi \bigg( e^{X}-\textbf{1}\bigg)_{ij}\simeq \psi \bigg(X_{ij}+\frac12X_{ik}X_{kj}\bigg).
\ee
The field strength tensor generated by the gauge field \eqref{A-tensor} to order two is
\bse\label{dFmunu}
\begin{align}\label{F0i}
F^a_{~0i}&=\delta^{aj}\bigg[\partial_t\!(a\psi)\delta_{jk}+a\psi\partial_t\!(X_{jk}+\frac12\gamma_{jk})\bigg](e^{X+\frac12\gamma})_{ki},\\
\label{Fij}
 F^a_{~ij}&=2a\psi\delta^{ak}(e^{X+\frac{\gamma}{2}})_{l[j}\partial_{i]}\bigg(X_{lk}+\frac12\gamma_{lk}\bigg)+\g a^2\psi^2\epsilon^{akl}(e^{X+\frac{\gamma}{2}})_{ki}(e^{X+\frac{\gamma}{2}})_{lj}.
\end{align}
\ese
while $F_a^{~0i}$ and $F_a^{~ij}$ are given as
\bse
\begin{align}\label{Fu0i}
F_a^{~0i}&=-a^{-2}\delta^{aj}\bigg[\partial_t\!(a\psi)\delta_{jk}+a\psi\partial_t\!(X_{jk}+\frac12\gamma_{jk})\bigg](e^{X-\frac12\gamma})_{ki},\nonumber\\
F_a^{~ij}&=2a^{-3}\psi\delta^{ak}(e^{X-\frac{\gamma}{2}})_{l[j}(e^{-\gamma})_{i]m}\partial_{m}\bigg(X_{lk}+\frac12\gamma_{lk}\bigg)+\g a^{-2}\psi^2\epsilon^{akl}(e^{X-\frac{\gamma}{2}})_{ki}(e^{X-\frac{\gamma}{2}})_{lj}.\nonumber
\end{align}
\ese
Therefore, we have
\be
F_a^{~0i}F^a_{~0i}=-a^{-2}\bigg[\partial_t\!(a\psi)\delta_{jk}+a\psi\partial_t\!(X_{jk}+\frac12\gamma_{jk})\bigg]\bigg[\partial_t\!(a\psi)\delta_{jm}+a\psi\partial_t\!(X_{jm}+\frac12\gamma_{jm})\bigg](e^{2X})_{km},\nonumber
\ee
which after using \eqref{Xtog}, we can read the second order $F_a^{~0i}F^a_{~0i}$ as
\be
\dtwo( F_a^{~0i}F^a_{~0i})=-a^{-2}\bigg([\partial_{t}\!(a\tg_{ji})]^2+\frac{a^2\psi^2}{4}(\dot\gamma_{ij})^2+a\partial_t\!(a\psi)\tg_{ij}\dot\gamma_{ij}+a^2\psi\dot{\gamma}_{ij}\dot{\tg}_{ij}\bigg).
\ee
The other term in the Yang-Mills theory, $F_a^{~ij}F^a_{~ij}$, is
\bea
&&F_a^{~ij}F^a_{~ij}=g^2\psi^4\bigg[\bigg(\textmd{tr}[e^{2X}]\bigg)^2-\textmd{tr}[e^{4X}]\bigg]+2a^{-2}\psi^2\partial_i\bigg(X_{lk}+\frac12\gamma_{lk}\bigg)\partial_j\bigg(X_{mk}+\frac12\gamma_{mk}\bigg)\times\nonumber\\
&&\bigg[(e^{2X})_{ml}(e^{-\gamma})_{ij}-(e^{X-\frac{\gamma}{2}})_{mi}(e^{X})_{jl}\bigg]+4\g\psi^3a^{-1}\epsilon^{ijk}(e^{X-\frac12\gamma})_{jm}(e^{2X})_{kn}\partial_m\bigg(X_{in}+\frac12\gamma_{in}\bigg),\nonumber
\eea
which at the second order 
and after using \eqref{Xtog} is given as
\be
\dtwo( F_a^{~ij}F^a_{~ij})=2a^{-2}[\partial_k (\tg_{ij}+\frac12\psi\gamma_{ij})]^2+4\g\psi a^{-1}\epsilon^{ijk}\tg_{kl}\partial_j\bigg(\tg_{il}+\frac{\psi}{2}\gamma_{il}\bigg).
\ee
Therefore the Yang-Mills has the following contribution to the quadratic action of tensor perturbations 
\bea\label{YM-T}
&&-\frac{\sqrt{-g}}{4}\dtwo(F^a_{~\mu\nu}F_a^{~\mu\nu})=\frac{a^3}{2}\bigg(\dot{\tg}_{ij}^2-a^{-2}(\partial_k \tg_{ij})^2-(\dot H+2H^2)\tg_{ij}^2
+2\g\psi a^{-1}\epsilon^{ijk}\tg_{il}\partial_j\big(\tg_{kl}+\frac{\psi}{2}\gamma_{kl}\big)\nonumber\\
&&+\psi\dot{\gamma}_{ij}\dot{\tg}_{ij}-\psi a^{-2}\partial_k\tg_{ij}\partial_k\gamma_{ij}
+a\partial_{t}\!(a\psi)\tg_{ij}\dot{\gamma}_{ij}+\frac14\psi^2 (\dot{\gamma}_{ij}^2-\partial_k\gamma_{ij}^2)\bigg).
\eea

Moreover, writing the Chern-Pontryagin density in terms of the Chern-Simons current as
\bea
\frac12F^a_{\mu\nu}\tilde F_a^{\mu\nu}=\epsilon^{\mu\nu\lambda\sigma}\partial_{\mu}\bigg(A^a_{~\nu}\partial_{\lambda}A^a_{~\sigma}+\frac13\g\epsilon^{abc}A^a_{~\nu}A^b_{~\lambda}A^c_{~\sigma}\bigg),
\eea
we arrive at
\be\label{CS_}
\frac12\varphi F^a_{\mu\nu}\tilde F_a^{\mu\nu}=-\dot{\varphi}a^{-3}\epsilon^{ijk}\bigg(A^a_{~i}\partial_{j}A^a_{~k}+\frac13\g\epsilon^{abc}A^a_{~i}A^b_{~j}A^c_{~k}\bigg).
\ee
For our convenience, here we introduce $$Y_{ij}\equiv X_{ij}+\frac12\gamma_{ij},$$ where $\textmd{tr}(\gamma)=0$ and $X_{ij}+\frac12X_{ij}^2\equiv\tg_{ij}$.
In terms of $Y_{ij}$, we can write the last term in \eqref{CS_} as
\bea
&&a^{-3}\epsilon^{ijk}\epsilon^{abc}\bigg(A^a_{~i}A^b_{~j}A^c_{~k}\bigg)=\psi^3\epsilon^{ijk}\epsilon^{lmn}\bigg((e^{Y})_{il}(e^{Y})_{jm}(e^{Y})_{kn}\bigg)=3!\psi^3\det(e^{Y})\nonumber\\
&&=3!\psi^3e^{\textmd{tr}(Y)}=3!\psi^3e^{\textmd{tr}(X)},
\eea
which in terms of $\tg_{ij}$ in \eqref{Xtog}, we can write it as
\bea
a^{-3}\epsilon^{ijk}\epsilon^{abc}\bigg(A^a_{~i}A^b_{~j}A^c_{~k}\bigg)=3!\psi\bigg(\psi^2-\frac{1}{2}\tg_{ij}^2 \bigg).
\eea
That then gives its second order of perturbation as
\be\label{CSt1}
\frac13a^{-3}\g\epsilon^{ijk}\epsilon^{abc}\dtwo\bigg(A^a_{~i}A^b_{~j}A^c_{~k}\bigg)=-\g\psi\tg_{ij}^2.
\ee
It is straightforward to see that
\be\label{CSt2}
\dtwo(\epsilon^{ijk}A^a_{~i}\partial_{j}A^a_{~k})=a^2\psi^2\epsilon^{ijk}Y_{il}\partial_{j}Y_{kl}+\textmd{total derivative}.
\ee
Finally, from \eqref{CSt1} and \eqref{CSt2} we arrive at
\bea\label{CS-T}
-\frac{\lambda\sqrt{-g}}{4f}\varphi\dtwo( F^a_{\mu\nu}\tilde F_a^{\mu\nu})=\frac{a^3}{2}\bigg( \frac{\dot{\varphi}\lambda}{af}\epsilon^{ijk}(\tg_{il}+\frac{\psi}{2}\gamma_{il})\partial_{j}(\tg_{lk}+\frac{\psi}{2}\gamma_{lk})-\frac{\dot{\varphi}\g\psi}{f}\tg_{ij}^2\bigg).
\eea

The total quadratic action for the spin-2 fluctuations is given as
\be\label{2nd-tot}
\dtwo S_{T}=\dtwo S_{G}+\dtwo S_{A},
\ee
where $\dtwo S_{G}$ and $\dtwo S_{A}$ are the contributions of perturbed gravitational sector and the gauge field sector respectively.
The perturbed Einstein-Hilbert action leads to
\be
\label{2ndR}
\dtwo S_{G}=\frac{1}{2}\int a^3d^3xdt \frac{1}{4}\bigg[\dot{\gamma}_{ij}^2-a^{-2}(\partial_k\gamma_{ij})^2\bigg],
\ee
while from the combination of \eqref{YM-T} and \eqref{CS-T}, we can find $\dtwo S_{A}$ as
\bea
\label{2ndLA}
&&\dtwo S_{A}=\frac{1}{2}\int a^3d^3xdt  \biggl[\dot{\tg}_{ij}^2-a^{-2}(\partial_k \tg_{ij})^2-(\dot H+2H^2+\frac{\lambda\dot{\varphi}\g\psi}{f})\tg_{ij}^2
+2(\frac{\lambda\dot{\varphi}}{2f}+\g\psi)a^{-1}\epsilon^{ijk}\tg_{il}\partial_{j}\tg_{lk}\nonumber\\
&&~~+\psi\dot{\gamma}_{ij}\dot{\tg}_{ij}-\frac{\psi}{a^2}\partial_k\tg_{ij}\partial_k\gamma_{ij}+\g\psi^2 a^{-1}\epsilon^{ijk}\tg_{il}\partial_j\gamma_{kl}
+\frac{\partial_{t}\!(a\psi)}{a}\tg_{ij}\dot{\gamma}_{ij}+\frac{\lambda\dot{\varphi}\psi}{2af}\epsilon^{ijk}(\tg_{il}\partial_{j}\gamma_{lk}+\gamma_{il}\partial_{j}\tg_{lk})\nonumber\\
&&+\frac14\psi^2 \bigg(\dot{\gamma}_{ij}^2-a^{-2}(\partial_k\gamma_{ij})^2
+\frac{\lambda\dot{\varphi}}{af}\epsilon^{ijk}\gamma_{il}\partial_j\gamma_{lk}\bigg)\biggr].
\eea

From the action \eqref{2ndLA}, we can read the field equation of $\tg_{ij}$ as
\be\label{FEQ-tg}
\ddot{\tg}_{ij}+3H\dot{\tg}_{ij}-a^{-2}\partial^2\tg_{ij}+2(1+\xi\xi_{\psi}-\frac12\epsilon)H^2\tg_{ij}-2(\xi+\xi_{\psi})Ha^{-1}\epsilon^{ilk}\partial_{l}\tg_{jk}\simeq0,
\ee
in which we neglect the effect of slow-roll suppressed interaction with metric. Moreover, varying the action \eqref{2nd-tot} with respect to $\gamma_{ij}$ and after using \eqref{FEQ-tg}, we find the field equation of graviton as
\be
\ddot{\gamma}_{ij}+3H\dot{\gamma}_{ij}-a^{-2}\partial^2\gamma_{ij}+2\psi\bigg(\ddot{\tg}_{ij}+3H\dot{\tg}_{ij}-a^{-2}\partial^2\tg_{ij}+H\dot{\tg}_{ij}+(3-\epsilon)H^2\tg_{ij}-(\xi_{\psi}+2\xi)Ha^{-1}\epsilon^{ilk}\partial_l\tg_{jk}\bigg)\simeq0, \nonumber
\ee
which after using \eqref{FEQ-tg}, we arrive at
\be
\ddot{\gamma}_{ij}+3H\dot{\gamma}_{ij}-a^{-2}\partial^2\gamma_{ij}+2\psi\bigg(H\dot{\tg}_{ij}+(1-2\xi\xi_{\psi})H^2\tg_{ij}+\xi_{\psi}Ha^{-1}\epsilon^{ilk}\partial_l\tg_{jk}\bigg)\simeq0.
\ee
From the above, we can read the anisotropic stress generated by the gauge field as
\be\label{piT-App}
\pi^T_{~ij}=H\psi\bigg((2\xi_{\psi}\xi-1)H\tg_{ij}-\dot{\tg}_{ij}
-\xi_{\psi}a^{-1}\epsilon^{ilk}\partial_l\tg_{jk}\bigg)\,.
\ee
As we see, the spin-2 perturbations of the SU(2) gauge field provides a source term for the gravity waves.

\section{Brief Overview of Bianchi Family in GR}\label{Bianchi-app}

The simplest generalization of the FRW model with six Killing vector fields are 
Bianchi cosmological models with three Killing vector fields. 
The Bianchi family are geometries with spatially homogeneous (constant $t$) surfaces which are invariant under the action of a three dimensional 
symmetry group, named after the classification scheme for 3-parameter Lie groups \cite{Bianchi}. The Bianchi space-time can be foliated into the spatial homogeneous hypersurfaces $\Sigma_t$,
$$M=\mathbb{R}\times\Sigma_t,$$
where $\mathbb{R}$ is the time variable. In each of these spatially homogeneous surfaces there exist a set of basis vectors $\textbf{e}_i$ that spans a Lie algebra as\footnote{The three Killing vectors, $X_i$, corresponding to $\{\textbf{e}_i\}$ satisfy $[\e_i,X_j]=0$ and $[X_i,X_j]=-C^k_{ij} X_k $.} 
\be\label{basis}
[\e_i,\e_j]=C^k_{~ij}\e_k.
\ee
Using the hypersurface normal one-form $\textbf{n}=dt$ and $\e^i$s which are the one-form basis duals to $\e_i$s ($\e^i.\e_j=\delta^i_j$), we can always write a Bianchi metric as (e.g. see \cite{Stephani:2003tm})
\be\label{metr}
ds^2=-dt^2+e^{2\alpha(t)}e^{2\beta_{ij}(t)}\e^i\otimes \e^j,
\ee
where $i,j=1,2,3$ label the coordinates in homogeneous space-like hypersurfaces, $e^{\alpha}$ is the isotropic scale factor and $\beta_{ij}$ is a traceless matrix which parametrizes the anisotropies. Bianchi classification categorizes the family of three dimensional Lie algebras into 9 different classes in which each algebra is labeled by a number I-IX.


Time evolution preserved the Bianchi type. Some types, however, permit isotropic sub-cases. Among the 9 different types of Bianchi models, some of them contain the standard FRW Universe. In particular, the flat FRW $(\rm k=0)$ is a special case of Bianchi types $\rm I$ and $\rm VII_0$, while the open FRW $(\rm k=-1)$ belongs to type $\rm V$ or $\rm VII_h$ and Bianchi type $\rm IX$ contains the closed FRW model $(\rm k=1)$. 
Here we focus on the Bianchi types I and $\rm VII_0$ which have flat space FRW cases.

\hskip 0.1 cm\textit{Bianchi type I:} In the Bianchi type I which is the simplest case, the Lie algebra is Abelian
\be
[X_i,X_j]=0,
\ee
which implies that the Killing vectors $X_i$, are the group of Galilean translations along the Cartesian coordinates $x^i$
\be
X_i=\partial_i.
\ee
Hence, we have $\e^1=dx$, $\e^2=dy$ and $\e^3=dz$.
Using the above and \eqref{metr}, we find the explicit form of the Bianchi type I metric as
\be\label{metr-B1}
ds^2=-dt^2+a_1^2(t)dx^2+a_2^2(t)dy^2+a_3^2(t)dz^2,
\ee
which has three different scale factors for each spatial direction.

\begin{figure*}[t] 
\begin{center}
\includegraphics[width=0.7\textwidth]{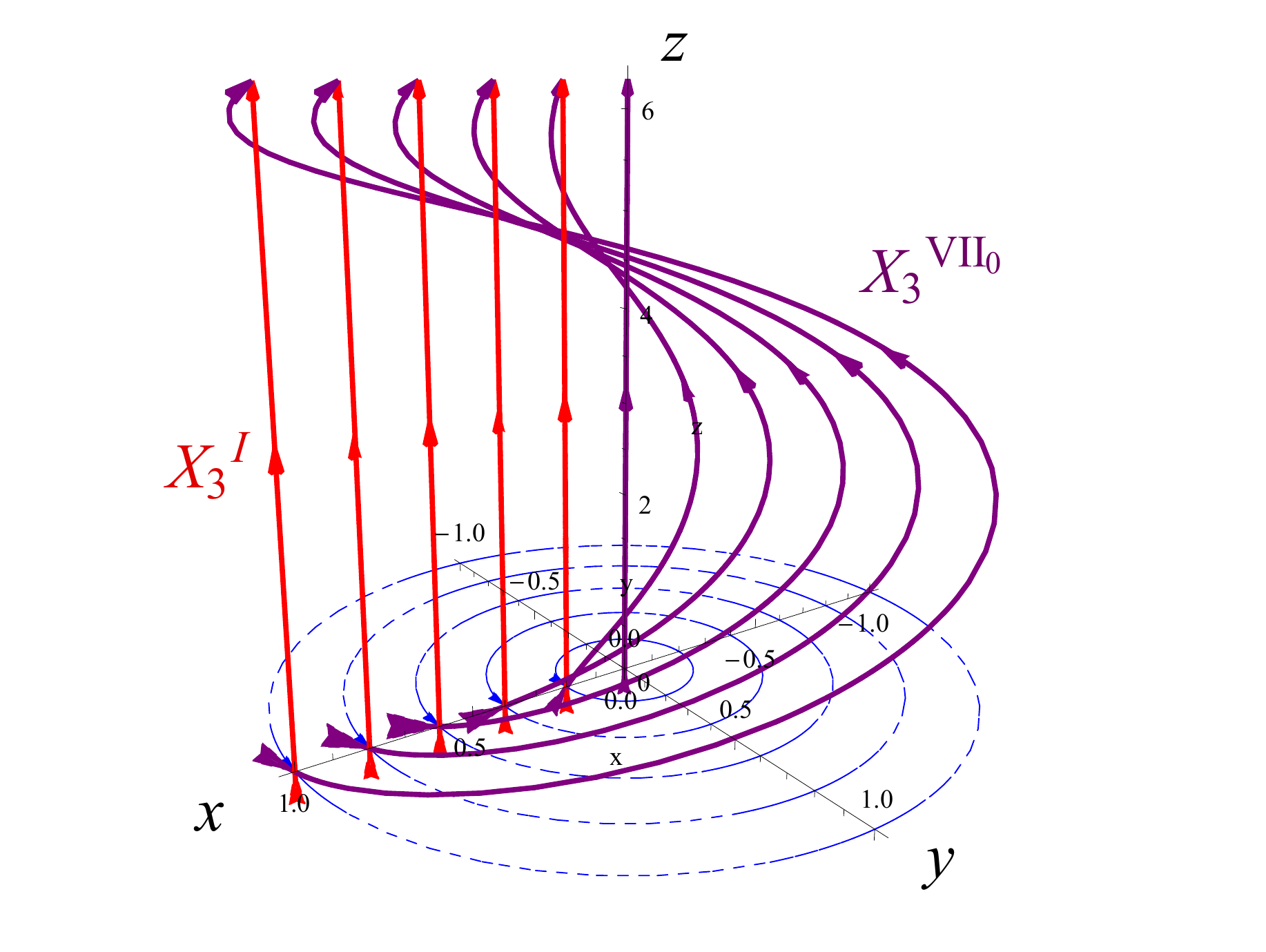}\\
\caption{The $X_3$ Killing vector field and its integral curves for Bianchi types $I$ and VII$_0$. The $X^{I}_3$ is equal to $T_3=\partial_z$ and induces a transformation along the z-axis. on the other hand, $X^{VII_0}_3=T_3+R_3$ and induces a complex spiralling motion along the z-axis. The solid purple lines are the integral curves of $X^{VII_0}_3$, the solid red lines show the integral curves of $X^{I}_3$ and the dashed blue lines represent $R_3$ respectively.}\label{fig-VII0-X3}
\end{center}  
\end{figure*}

\hskip 0.1cm\textit{Bianchi type $\rm VII_0$:} In the Bianchi type $\rm VII_0$, the non-zero structure constants are of the form $C_{13}^{2}=-C_{23}^1=-1$. That then leads to the following commutation relations for the Killing vectors
\bse
\begin{align}
&[X_1,X_2]=0,\\
&[X_1,X_3]=X_2,\\
&[X_2,X_3]=-X_1, %
\end{align}
\ese
which implies that $X_{1}$ and $X_{2}$ are the Galilean translations along $x$ and $y$ while $X_{3}$ is 
\be
X_3=T_3+R_3,
\ee
which is a combination of the translation along $z$ and the rotation around $z$-axis ($R_i=\epsilon_{ijk}x^j\partial_k$). That Killing vector field generates a spiralling motion along the axis $z$ (see fig. 2). Using $[X_i,\e_j]=0$, we find the basis vectors as\footnote{In fact $\e_a(z)=R_a^i(-z\hat{z})\partial_i$, where $R(\hat{z})$ is the rotation matrix around the $z$-axis.}
\bse\label{dxi-VII}
\begin{align}
\e_1(z)&=\cos z\partial_x+\sin z\partial_y,\\
\e_2(z)&=-\sin z\partial_x+\cos z\partial_y,\\
\e_3(z)&=\partial_3.
\end{align}
\ese
The one-form basis duals can be find using $\e^i.\e_j=\delta^i_j$. Finally, using \eqref{metr} and the above, we arrive at the following general form for a Bianchi type $VII_0$ metric
\be
ds^2=-dt^2+e^{2\alpha(t)}\big[ e^{2\tilde\beta(t)}dz^2+e^{2\beta_{\rm ab}(t)}\e^{\rm a}\otimes\e^{\rm b}\big],
\ee
where $\rm a, b=1,2$, $\alpha(t)$ and $\beta(t)$ are two functions of time while $\beta_{\rm ab}(t)$ is a time-dependent $2\times2$ traceless matrix. In the coordinate basis, we can write the above metric as
\be
ds^2=-dt^2+e^{2\alpha(t)}\big[dx^2+dy^2+e^{2\tilde\beta(t)}dz^2+\gamma_{ ij}(t,\textbf{x})dx^{ i}dx^{j}\big],
\ee
where $\gamma_{\rm ij}(t,\textbf{x})$ is a traceless tensor given as
\be\label{Bianchi-VII0}
\gamma_{ ij}(t,\textbf{x})=\left(
\begin{tabular}{ccc}
$\cos(\textbf{q}.\textbf{x})$ & $\sin(\textbf{q}.\textbf{x})$ & 0 \\
$\sin(\textbf{q}.\textbf{x})$ & -$\cos(\textbf{q}.\textbf{x})$ & 0  \\
0 & 0 & 0\\
\end{tabular}
\right),
\ee
in which \textbf{q} is directed along the $z$-axis and its magnitude $q$ is an arbitrary constant corresponding to the scale transformation of the $z$-axis. It is interesting to notice that while the Bianchi metric is only time-dependent in the one-form basis, it can be a function of space-time in the coordinate basis. The above $\gamma_{ ij}(t,\textbf{x})$ tensor is a standing circularly polarized gravitational wave and the positive (negative) value of $q$ corresponds to a helicity $+2$ ($-2$) state of the gravitational wave.

{\footnotesize\bibliography{ref}}

\end{document}